\documentclass[11pt]{article}
\usepackage{jcappub}
\usepackage{array, natbib, amsfonts}
\usepackage{amssymb, amsmath}
\usepackage{epsfig}
\usepackage{graphicx,color}
\usepackage{placeins}

\newcommand{\hMpc}{{h^{-1}\,{\rm Mpc}}}
\newcommand{\hMpcinv}{{h\,{\rm Mpc}^{-1}}}

\newcommand{\kmax}{{k_{\rm max}}}

\newcommand{\ah}{{\alpha_H}}
\newcommand{\ad}{{\alpha_D}}
\newcommand{\gh}{{\gamma_H}}
\newcommand{\gd}{{\gamma_D}}

\newcommand{\mnras}{{\rm MNRAS}}
\newcommand{\apj}{{\rm ApJ}}
\newcommand{\apjs}{{\rm ApJS}}

\newcommand{\jcap}{{\rm JCAP}}

\newcommand{\nat}{{\rm Nature}}
\newcommand{\prd}{{\rm Phys.Rev.D}}
\newcommand{\apjl}{{\rm ApJL}}

\newcommand{\physrep}{{\rm Phys. Rep.}}

\newcommand{\plotdir}{Figures/}

\title{Geometric and dynamic distortions in anisotropic galaxy clustering}

\author[a,b]{Jonathan Blazek,}
\author[a,c,d]{Uro\v{s} Seljak,}
\author[c]{Zvonimir Vlah,}
\author[d]{and Teppei Okumura}
\affiliation[a]{Departments of Physics and Astronomy, and Lawrence Berkeley National Laboratory,\newline University of California, Berkeley, CA 94720, USA}
\affiliation[b]{Center for Cosmology and AstroParticle Physics, The Ohio State University,\newline Columbus, OH 43210, USA}
\affiliation[c]{Institute of Theoretical Physics, University of Zurich, CH-8057, Zurich, Switzerland}
\affiliation[d]{Institute for the Early Universe, Ewha University, Seoul 120-750, South Korea}
\emailAdd{blazek@berkeley.edu}

\abstract{We examine the signature of dynamic (redshift-space) distortions and geometric distortions (including the Alcock-Paczynski effect) in the context of the galaxy power spectrum measured in upcoming galaxy redshift surveys. Information comes from both the baryon acoustic oscillation (BAO) feature and the broadband power spectrum shape. Accurate modeling is required to extract this information without systematically biasing the result. We consider an analytic model for the power spectrum of dark matter halos in redshift space, based on the distribution function expansion, and compare with halo clustering measured in N-body simulations. We forecast that the distribution function model is sufficiently accurate to allow the inclusion of broadband information on scales down to $k \sim 0.2 \hMpcinv$, with somewhat better accuracy for higher bias halos. Compared with a BAO-only analysis with reconstruction, including broadband shape information can improve unbiased constraints on distance measures $H(z)$ and $D_A(z)$ by $\sim 30\%$ and $20\%$, respectively, for a galaxy sample similar to the DESI luminous red galaxies. The gains in precision are larger in the absence of BAO reconstruction. Furthermore, including broadband shape information allows the measurement of structure growth, through redshift-space distortions. For the same galaxy sample, the distribution function model is able to constrain $f \sigma_8$ to  $\sim2\%$, when simultaneously fitting for $H(z)$ and $D_A(z)$. We discuss techniques to optimize the analysis of the power spectrum, including removing modes near the line-of-sight that are particularly challenging to model, and whether these approaches can improve parameter constraints. We find that such techniques are unlikely to significantly improve constraints on geometry, although they may allow higher precision measurements of redshift-space distortions.
}
\arxivnumber{}

\begin{document}
\maketitle
\flushbottom

\section{Introduction}

Galaxy redshift surveys are a primary tool in probing the universe, including the nature of dark matter and dark energy. The baryon acoustic oscillation (BAO) feature (see, e.g.,~\cite{eisenstein98}) can be used as a ``standard ruler'' to measure geometry and the expansion history of the universe. Beyond this feature, the full scale-dependence of galaxy clustering can test a number of important components of the cosmological model, including the epoch of matter-radiation equality, neutrino mass, non-Gaussianity in the initial density fluctuations, and nonlinear structure growth (e.g.\ \cite{tegmark04,lesgourgues06,dalal08}).

Galaxy redshift surveys are typically used to create a ``map'' in real space of the three-dimensional positions of galaxies. However, these surveys actually measure two angular coordinates and a redshift for each object, which must be converted into three-dimensional positions assuming an underlying geometry of the expanding universe. This conversion, specified by the Hubble parameter $H(z)$ and the angular-diameter distance $D_A(z)$, will introduce distortions into the resulting galaxy map if an incorrect geometry is assumed. Comparing an observed feature to a known physical scale can probe this geometry. Similarly, in an isotropic universe, clustering strength should not depend on orientation with respect to the observer. However, an anisotropic signal can arise since separations along the line-of-sight are measured differently than those perpendicular to it. This effect is a version of the Alcock-Paczynski (AP) test \citep{alcock79}, a method to measure the expansion of the universe using a spherically symmetric object (or feature), even without a known physical scale. The observed galaxy clustering signal will manifest these geometric distortions through both shifting of known physical scales and a warping of the initially isotropic clustering strength. While a fully general analysis would treat these effects together, many past studies have focused on the shifting of the BAO scale, which is considered more robust to modeling uncertainties \citep{eisenstein07} but provides no constraint on structure growth via redshift-space distortions.

Galaxies, or other objects being mapped in redshift surveys, also have peculiar velocities which introduce dynamical distortions to the observed clustering signal. The line-of-sight component of peculiar velocity contributes to the observed redshift and is thus degenerate with the cosmological redshift. This effect is commonly known as ``redshift-space distortions'' (RSD). RSD contain information about velocity fields and can thus provide a powerful probe of the growth of structure and potential modifications to general relativity (e.g.\ \cite{jennings11b}). Indeed, measurements of RSD have the potential to be one of the primary sources of cosmological information in upcoming surveys (e.g.\ \cite{font13}). However, since the signatures of geometric and dynamical distortions can be quite similar, a reliable understanding of the latter is necessary to separate the two and interpret the observed clustering signal in a cosmological context. While a linear theory description of galaxy clustering in redshift space has long been available \citep{kaiser87}, significant efforts have been made in recent years to improve our analytic understanding of clustering and redshift-space distortions in the nonlinear regime (e.g.\ \cite{scoccimarro04,taruya10,seljak11,vlah12,vlah13,okumura12b,okumura12a,vallinotto13}). 

Most prior attempts to constrain geometry from measurements of galaxy clustering have focused on measuring the angle-averaged BAO feature (e.g.\ \cite{eisenstein05,percival10,blake11a,anderson12}), which constrains a combination of distance scales approximated as $D_V(z) = D_A(z)^2/H(z)$. The greater statistical power of recent surveys has allowed the use of anisotropic clustering information from the BAO feature, the broadband clustering shape, or both to measure $D_A(z)$ and $H(z)$ separately, as well as constraining redshift-space distortions \citep{okumura08,blake11b,reid12,anderson13,kazin13,sanchez13,beutler13}. Current and planned redshift surveys, including BOSS,\footnote{Baryon Oscillation Spectroscopic Survey; http://www.sdss3.org/surveys/boss.php}
eBOSS,\footnote{Extended Baryon Oscillation Spectroscopic Survey; http://www.sdss3.org/future/eboss.php}
DESI,\footnote{Dark Energy Spectroscopic Instrument \citep{levi13}}
and EUCLID,\footnote{http://sci.esa.int/science-e/www/area/index.cfm?fareaid=102}
offer the opportunity to probe these geometric and dynamical effects at a high level of statistical precision, requiring a thorough understanding of model uncertainties and how to extract cosmological information from the clustering signal.

Previous studies have established the theoretical framework for measuring geometric and dynamical information from galaxy clustering \citep{ballinger96,padmanabhan08b,montanari12,kwan12,linder13} and have examined the constraining power and modeling requirements of such an analysis \citep{shoji09,taruya11,kazin12,gilmarin12,song13,oka13}. In this work, we discuss the information content in the full anisotropic galaxy clustering signal and compare with that in the BAO feature alone. While most previous studies focus on models of dark matter clustering, we consider the clustering of the dark matter halos in which galaxies reside. Challenges in modeling anisotropic clustering in redshift space, particularly on small scales and for separations along the line-of-sight, introduce systematic biases in cosmological parameter estimates. In light of these biases, we employ Fisher matrix formalism to determine the minimum scales that can be reliably used and discuss techniques to optimize the analysis. Several analytic models exist for galaxy clustering in redshift space. We focus in particular on the recently developed distribution function approach \citep{seljak11}, which provides an accurate description down to comparatively small scales. While we work with the galaxy power spectrum, many recent measurements of anisotropic galaxy clustering have been done using the correlation function, for which analogous arguments apply.

This paper is organized as follows. In Section \ref{sec:formalism}, we summarize the general formalism for galaxy clustering in redshift space, including the effect of geometric distortions, and describe the N-body simulations used to test analytic models. In Section~\ref{sec:DF}, we describe the distribution function approach and construct a model for the full shape of the halo power spectrum in redshift space. Section~\ref{sec:fisher} develops the Fisher matrix formalism and other elements of how we forecast the performance of clustering models, and Section~\ref{sec_GDD:results} presents the results. We conclude in Section~\ref{sec:discussion} with a summary and discussion of the major results. We assume a flat, $\Lambda$CDM fiducial cosmology with $\Omega_{\rm m}=0.279$, $\Omega_{\rm b}/\Omega_{\rm m}=0.165$, $h=0.701$, $\sigma_8=0.807$, and $n_{\rm s}=0.96$.

\section{Modeling galaxy clustering in redshift space}
\label{sec:formalism}
Numerous efforts have been made in recent years to model the clustering of galaxies in redshift space, using both N-body simulations (e.g.\ \cite{jennings11, okumura12b}) and analytic techniques (e.g.\ \cite{scoccimarro04, reid11}). Here we summarize the relevant aspects of the field and develop a useful expansion for the geometric distortions we wish to measure.

\subsection{Clustering in real and redshift space}
\label{sec:formalism1}
In real space, the power spectrum of density fluctuations $\delta$, or equivalently the two-point correlation function, depends only on the amplitude of the scale being considered:
\begin{align}
\label{eq:Pk_defined}
\langle \delta(\mathbf{k}) \delta^{*}(\mathbf{k'})\rangle = (2\pi)^3 \delta(\mathbf{k}-\mathbf{k'}) P(k),
\end{align}
where $k=|\mathbf{k'}|$. The density of a luminous tracer, such as a galaxy population, is related to that of dark matter through a biasing relationship. On large scales, a constant, linear biasing relationship is often assumed:
\begin{align}
\delta_{\rm gal} &= b_{1}\delta, \notag \\
P_{\rm gal}(k) &= b_{1}^2P_{\rm DM}(k).
\end{align}
However, this assumption breaks down on quasi-linear scales. Thus any model for galaxy clustering on small scales must consider not only the nonlinear clustering of dark matter, but also the complex bias relationship between dark matter and galaxies. In this work, we use the non-linear, non-local bias described in \cite{baldauf12}:
\begin{align}
\label{eq:nonlinear_bias}
\delta_{\rm gal} (\mathbf{x}) &= b_1\delta(\mathbf{x}) + \frac{b_2}{2} \left(\delta^2(\mathbf{x}) -\langle \delta^2\rangle \right) + \frac{b_s}{2} \left(s^2(\mathbf{x}) -\langle s^2\rangle \right).
\end{align}
Non-locality comes from the tidal term $s^2(\mathbf{x})=s_{ij}(\mathbf{x})s_{ij}(\mathbf{x})$, for tidal tensor $s_{ij}$:
\begin{align}
s_{ij}(\mathbf{x}) = \partial_i\partial_j\Phi(\mathbf{x}) - \frac{1}{3}\delta^{\rm K}_{ij}\delta(\mathbf{x}),
\end{align}
where $\delta^{\rm K}_{ij}$ is the Kronecker delta function. A local, third-order bias $b_3$ can be trivially absorbed into the value of $b_1$. The effect of a non-local third-order bias $b_3^{\rm NL}$ is discussed in Section~\ref{sec:biasing}.

Redshift-space distortions break the natural isotropy of Equation~\ref{eq:Pk_defined}, leading to a dependence on the angle with respect to the line-of-sight. The observed wavevector $\mathbf{k}$ can be decomposed into $(k_{\parallel},k_{\perp})$, parallel and perpendicular to the line-of-sight, respectively.\footnote{Rotational symmetry remains on the plane perpendicular to the line-of-sight.} Equivalently, one can use $k$ and $\mu=\cos\theta$, where $\theta$ is the angle between $\mathbf{k}$ and the \mbox{line-of-sight:}
\begin{align}
k_{\parallel}&=\mu k , \notag \\
k_{\perp}&=(1-\mu^2)^{1/2} k .
\end{align}
On large scales, in the plane-parallel approximation and with linear bias, redshift-space distortions are described by the Kaiser formula \citep{kaiser87}:
\begin{align}
\label{eq:kaiser}
P_{\rm gal}^{\rm s}(k,\mu)=P^{\rm r}_{\rm DM}(k)(b_1+f\mu^2)^2,
\end{align}
where ``s'' denotes redshift space, ``r'' denotes real space, and the logarithmic growth rate $f=d\ln G(a)/ d\ln a$, for scale factor $a$ and growth factor $G(a)$. In general relativity, $f\approx \Omega_m^\gamma$ with $\gamma \approx 0.55$, while theories of modified gravity can yield different values of $\gamma$ \citep{linder05}. Thus, a precise measurement of $f$ from redshift-space distortions can test gravitational physics. The relationship between density and velocity fields that yield Equation \ref{eq:kaiser} is only valid in the linear regime. Extending models of RSD to smaller scales requires a more detailed treatment of this relationship. A frequently used model is the ansatz of \cite{scoccimarro04} (hereafter the ``S04 model''), which includes the characteristic ``fingers-of-God'' (FoG) effect, in which the galaxy velocity dispersion $\sigma_v$ suppresses power on small scales:
\begin{align}
\label{eq:scoccimarro}
P_{\rm gal}^{\rm s}(k,\mu)=\left(b^2 P_{\delta \delta}(k) + 2bf\mu^2 P_{\delta \theta}(k) + f^2\mu^4 P_{\theta \theta}(k)\right)e^{-\left(f \sigma_v k \mu\right)^2},
\end{align}
where $P_{\delta\delta}$, $P_{\delta\theta}$, and $P_{\theta\theta}$ are the non-linear auto- and cross-power spectra of mass density and velocity divergence.\footnote{This ansatz was originally proposed as a simple model for the redshift-space power spectrum of matter alone, but it is sometimes used, in combination with a linear bias factor, to describe galaxies.}
\subsection{Parametrizing angular dependence}

$P(k,\mu)$ must be specified on the entire two-dimensional $k-\mu$ plane (or equivalently the $k_{\parallel}-k_{\perp}$ plane). By symmetry, the auto-power spectrum in redshift space can be expanded in even powers of $\mu$:
\begin{align}
\label{eq:muexpansion}
P^{\rm s}(k,\mu)=\sum_{j=0} F_{2j}(k)\mu^{2j}.
\end{align}
This expansion should be convergent on sufficiently large scales, and the maximum $j$ required to accurately describe the angular dependence increases with the maximum wavenumber considered, $k_{\rm max}$. Similarly, it is common to express the angular dependence in terms of a multipole expansion:
\begin{align}
\label{eq:multipoleexpansion}
P^{\rm s}(k,\mu)=\sum_{j=0} A_{2j}(k)\mathcal{P}_{l=2j}(\mu),
\end{align}
where $\mathcal{P}_{l}$ are the Legendre polynomials. The multipole expansion is particularly convenient from an observational perspective, since the orthogonality of the the Legendre polynomials yields a roughly diagonal covariance matrix.\footnote{The survey window function and anisotropic noise properties due to RSD can induce small off-diagonal covariance between multipoles.} Measurements of the angle-averaged power spectrum include only the monopole ($l=0$), while some recent studies (e.g.\ \cite{anderson13}) have also used the quadrupole ($l=2$). It was shown in \cite{taruya11} that including terms up to the hexadecapole ($l=4$) recovers most of the information contained in the full $P^{\rm s}(k,\mu)$. 

In this work, we use the full 2D clustering shape. As discussed below, we generate $P^{\rm s}(k,\mu)$ using an expansion in powers of $\mu$, truncated at $\mu^6$ in simulations and $\mu^4$ for analytic models. For the remainder of this work, the ``s'' superscript is omitted from $P(k,\mu)$, which is assumed to be in redshift space unless otherwise noted.

\subsection{Geometric distortions}
\label{sec:geo_dist}

If the fiducial (assumed) values of the angular diameter distance, $D_A(z)$, and the Hubble parameter, $H(z)$, differ from their true values, $k_{\parallel}$ and $k_{\perp}$ are affected:
\begin{align}
D_A(z)^{\rm true}&=\alpha_D^{-1} D_A(z)^{\rm fid}, \\
k_{\perp}^{\rm true}&=\alpha_D k_{\perp}^{\rm fid}, \notag\\
H(z)^{\rm true}&=\alpha_H H(z)^{\rm fid}, \notag\\
k_{\parallel}^{\rm true}&=\alpha_H k_{\parallel}^{\rm fid}, \notag
\end{align}
where the ``fid'' superscript indicates that the potentially incorrect fiducial cosmology has been applied. For simplicity, this superscript is dropped in the remainder of this work. One intuitive parametrization of these geometric deviations involves an isotropic ``dilation'': $\alpha= (\ad^2\ah)^{-1/3}$; and an anisotropic ``warping'': $\epsilon=(\ad/\ah)^{1/3}-1$ \citep{padmanabhan08b}. In the absence of redshift-space distortions (e.g.\ if they are removed through a reconstruction process), the position of a feature in the angle-averaged clustering signal measures $\alpha$, while the angular dependence of clustering measures $\epsilon$. We choose to use the direct $\ah$ and $\ad$ parametrization which is completely equivalent and may be easily applied to general clustering studies (e.g.\ the Lyman-$\alpha$ forest) where the quantities most directly measured do not correspond to $\alpha$ and $\epsilon$. Note that in the presence of RSD, the quantity measured by an angle-averaged BAO measurement from galaxies can also deviate from $\alpha$ (e.g.\ \cite{percival10}).

For convenience, we introduce the following small quantities:
\begin{align}
\gh=\ah^2-1\approx 2\left(\ah-1 \right), \\
\gd=1-\ad^2 \approx 2\left( 1-\ad \right). \notag
\end{align}
In terms of their fiducial values, the true wavevector amplitude and orientation are:
\begin{align}
k^2_{\rm true} &=k^2\left(1-\gd\left(1-\mu^2\right)+\gh\mu^2\right), \\
\mu^2_{\rm true}&=\mu^2\left(\frac{1+\gh}{1-\gd\left(1-\mu^2\right)+\gh\mu^2}\right). \notag
\end{align}

In addition to assigning the observed power spectrum to incorrect values of $k$ and $\mu$, geometric distortions also introduce a multiplicative correction, $\Delta V$, due to the difference in volume between the true and assumed cosmologies (e.g.\ \cite{ballinger96}):
\begin{align}
\Delta V \equiv \left. \left(\frac{D_A^2(z)}{H(z)}\right)\middle/\left(\frac{D_A^2(z)}{H(z)}\right)_{\rm true}\right. = \frac{D_V}{D_V^{\rm true}} = (1+\gh)^{1/2}(1-\gd).
\end{align}
Note that for the correlation function $\xi(\mathbf{r})$, which is dimensionless, the geometric distortions appear purely as rescalings of $(r_{\perp}, r_{\parallel})$, and there is no additional volume correction. Together, these effects yield:
\begin{align}
\label{eq:distort_general}
P_{\rm obs}(k,\mu)=\Delta V P_{\rm true}(k_{\rm true},\mu_{\rm true}).
\end{align}
As described above, we can generically write the true angle-dependent power spectrum:
\begin{align}
P_{\rm true}(k_{\rm true},\mu_{\rm true})=\sum_{j=0} F_{2j}(k_{\rm true})\mu_{\rm true}^{2j}.
\end{align}
Applying geometric distortions (i.e.\ Equation~\ref{eq:distort_general}) to this expansion gives:
\begin{align}
\label{eq:APexpansion1}
P_{\rm obs}(k,\mu)=(1+\gh)^{1/2}(1-\gd) & \sum_{j=0}\mu^{2j}\left(\frac{1+\gh}{1-\gd\left(1-\mu^2\right)+\gh\mu^2}\right)^j \notag \\
& \times F_{2j}\left(k\left(1-\gd\left(1-\mu^2\right)+\gh\mu^2\right)^{1/2}\right).
\end{align}
This expression can be expanded to arbitrary order in $\gh$ and $\gd$. When the distortions are small, the first-order expansion is sufficiently accurate: 
\begin{align}
\label{eq:AP_first_order}
P_{\rm obs}(k,\mu)&\approx\left(\frac{1}{2}\gh - \gd \right)P_{\rm true}(k,\mu) \\
&+\mu^0\left[ F_0^{(0)}-(k^2\gd)F_0^{(1)}\right] \notag\\
&+ \mu^2 \left[(1+\gd+\gh)F_1^{(0)} + k^2(\gd+\gh)F_0^{(1)} -(k^2\gd)F_1^{(1)}\right] \notag \\
&+ \mu^4 \left[(1+2\gd+2\gh) F_2^{(0)} -(\gd+\gh) F_1^{(0)}  +k^2(\gd+\gh) F_1^{(1)} -(k^2\gd) F_2^{(1)} \right] \notag \\
&+ \mu^6 \left[ (1+3\gd+3\gh)F_3^{(0)}-2(\gd+\gh) F_2^{(0)} +k^2(\gd+\gh) F_2^{(1)} -(k^2\gd) F_3^{(1)}\right] \notag \\
&+ \cdots , \notag
\end{align}
where $F_j^{(n)} \equiv (\partial /\partial k^2)^n F_j$, and terms scaling as $\mu^8$ and above aren't shown. The first line of the equality shows the overall amplitude shift due to the fractional change in volume, $\Delta V$. We use this expansion to include geometric distortions in the Fisher matrix formalism described in Section~\ref{sec:fisher}.

The shift of clustering power from lower to higher powers of $\mu$ is one signature of these geometric distortions. For instance, the Kaiser formula predicts zero power at $\mu^6$ and above (or the equivalent multipoles). In principle, measuring this angular dependence on large scales (where linear theory holds) would be a clean detection channel for geometric distortions. There are significantly more modes on small scales (high $k$), providing the possibility of much higher signal-to-noise, although using these smaller scales requires understanding nonlinear effects. Accurate modeling of nonlinear clustering and redshift-space distortions helps extract information in two ways. First, the broadband and BAO features in the power spectrum provide specific distance scales that can be used as standard rulers. Second, an understanding of how redshift-space distortions induce anisotropy in the power spectrum is required to perform an Alcock-Paczynski test.

\subsection{N-body simulations}
\label{sec:sim}
We use the clustering of dark matter halos in N-body simulations as a reference against which to compare different models. Since galaxies reside in dark matter halos, N-body simulations provide an important link in understanding observed clustering and redshift-space distortions. However, the physics of galaxy formation and existence of satellite galaxies can lead to a non-trivial relationship between halos and galaxies (e.g.\ \cite{reid09}). We consider the clustering of dark matter halos and leave the halo-galaxy relationship for future work (see, e.g., \cite{nishimichi13} for a recent exploration of this relationship in the context of galaxy clustering). The modeling of central galaxies, which should exhibit clustering properties similar to halos, is more straightforward. Some galaxy types used in clustering measurements, e.g.\ luminous red galaxies (LRGs), are primarily central objects, and it may be possible to construct a sample with a low level of satellite contamination \citep{reid09}.

The power spectra in the distribution function expansion, described in Section~\ref{sec:DF}, are from mass-weighted velocity moments and are thus straightforward to determine from simulations, since the contribution from empty grid cells is well-defined.  We use results calculated from $N$-body simulations as described in \cite{okumura12b,okumura12a}. We provide a brief summary of these simulations here.

We employ a series of $N$-body simulations of the $\Lambda$CDM cosmology seeded with Gaussian initial conditions, an updated version of \cite{desjacques09}.  The fiducial cosmology corresponds to the best-fit parameters in the WMAP 5-year data
\citep{komatsu09}, with $\Omega_{\rm m}=0.279$, $\Omega_{\rm b}=0.0462$, $h=0.701$, $n_{\rm s}=0.96$, and a
normalization of the curvature perturbations corresponding to a density fluctuation amplitude $\sigma_8 = 0.807$. A total of $1024^3$ particles of mass $m_p=2.95\times 10^{11} h^{-1}M_\odot$ are placed in a cubic box with side length $1600\hMpc$. To reduce sample variance, 12 simulations are used and each of the three lines-of-sight are treated as independent for 36 total realizations.

Dark matter halos are identified at the four redshifts using the
friends-of-friends algorithm \citep{davis85} with a linking length equal to 0.17 times
the mean particle separation. Halos must have at least 20 particles and are divided into subsamples by mass. Properties of the halo catalogs at $z=0$ and 0.509 (quoted as $z=0.5$) are summarized in Table \ref{tab:halo}. In most of this work, we focus on the two lowest halo mass bins at $z=0.5$.

\begin{table}[t!]
\begin{center}
\begin{tabular}{c | ccccccc}
$z$ & Mass & Mass range & $ \bar{N}$ & $\bar{n}$ & $b_1$ & $b_{2,00}$ & $b_{2,01}$ \\
        & bin & $(10^{12}h^{-1}M_\odot)$ & ($\times10^4$)  & $(h^3{\rm Mpc}^{-3})$ &(cross)& & \\
\noalign{\hrule height 1pt}
0 & $1$ & $5.91 - 17.7$ & 175 & $4.28\times 10^{-4}$ &$1.18$  & -0.39 & -0.45\\
& $2$ & $17.7-53.2$ & $63.3$ & $1.54\times 10^{-4}$  &$1.47$ & -0.08 & -0.35\\
&$3$ & $53.2-159$ & $18.7$ & $4.57\times 10^{-5}$& $2.04$ & 0.91 & 0.14\\
&$4$ & $159-467$ & $4.05$ & $9.89\times 10^{-6}$&$3.05$ & 3.88 & 2.0\\
\hline
0.5 &$1$ & $5.91 - 17.7$ & $144$ & $3.51\times 10^{-4}$&$1.64$& 0.18 & -0.20\\
&$2$& $17.7-53.2$ & $44.8$ & $1.09\times 10^{-4}$&$2.18$ & 1.29 & 0.48\\
&$3$& $53.2-159$ & $9.96$ & $2.43\times 10^{-5}$&$3.13$ & 4.48 & 2.6\\
&$4$& $159-467$ &  $1.30$ & $3.18\times 10^{-6}$& $4.82$ & 12.65 & 9.5\\
\end{tabular}
\end{center}
\caption{Properties of halo catalogs. $\bar{N}$ and $\bar{n}$ are the average number and number density of halos in each realization, respectively. The linear bias values $b_1$ are computed from the cross-power spectrum ($P^{mh}_{00}$), averaged at $0.01\leq k\leq 0.04 \hMpcinv$. The quadratic bias values $b_{2}^{00}$ and $b_{2}^{01}$ are fit to $P_{00}$ and $P_{01}$, respectively, as in \cite{vlah13}. The $b_2$ values shown here are fit after applying a correction to the relevant perturbation theory terms, as discussed in Section~\ref{sec:combining_terms}. Fitting without this correction yields different $b_2$ values.}
\label{tab:halo}
\end{table}

\begin{figure}[h!]
\begin{center}
\resizebox{\hsize}{!}{
\includegraphics{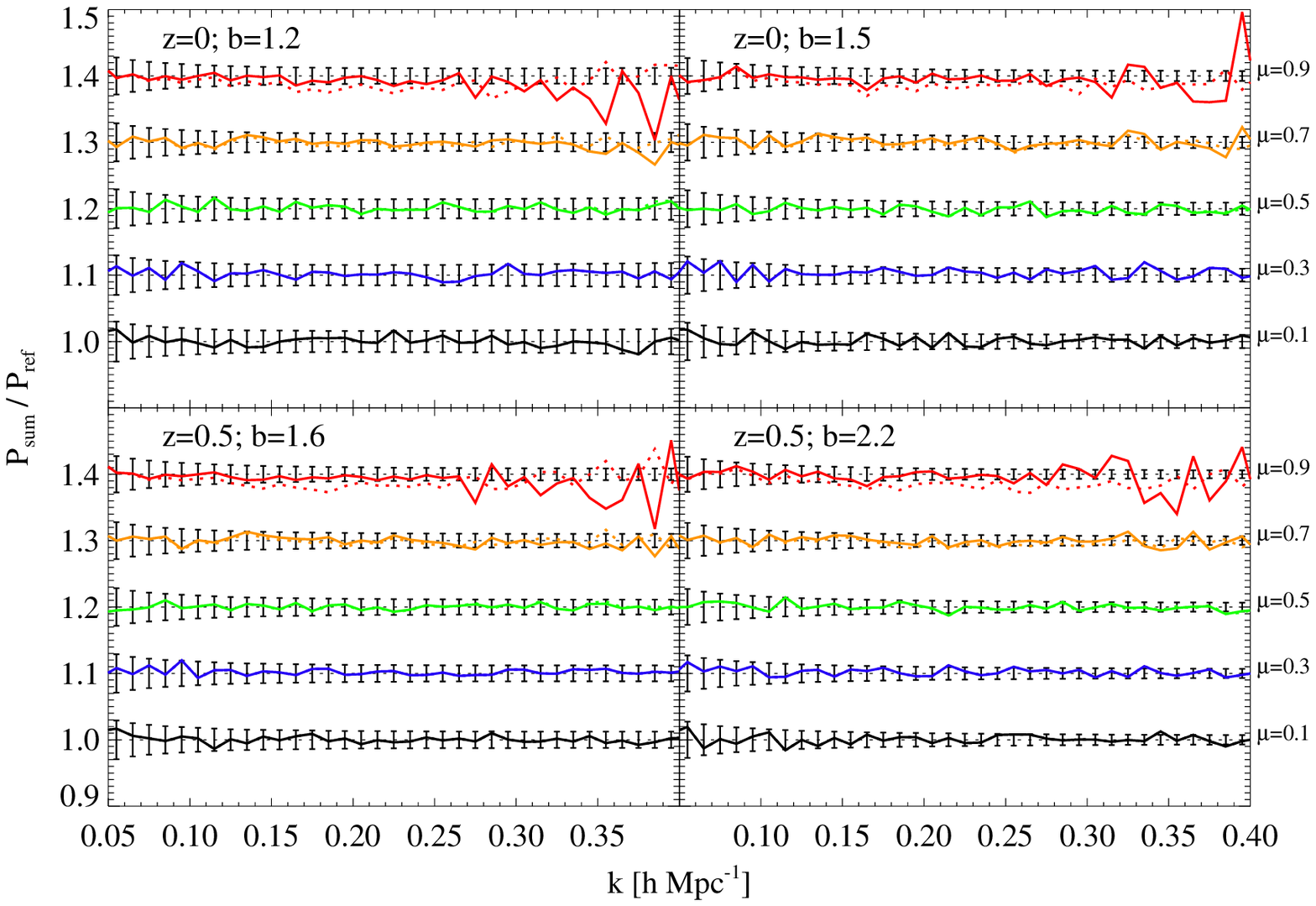}
}
\end{center}
\caption{$P(k,\mu)$ calculated by summing over $\mu^{2j}$ terms is shown, normalized by the reference $P(k,\mu)$ calculated directly from simulations. The full scale-dependent shot noise $\Lambda(k)$ is subtracted from both the reference and summed $P(k,\mu)$; see Equation \ref{eq:lambdak}. {\it Left panels} show the lowest halo mass bin, and {\it right panels} show the second mass bin; $z=0$ ({\it top panels}) and $z=0.5$ ({\it bottom panels}). See Table~\ref{tab:halo} for more information on halo mass and bias. Five evenly spaced angular bins for $0 < \mu< 1$ are shown, with vertical offsets added for clarity and central $\mu$ values as labelled. Error bars show the expected (fractional) uncertainty for a survey similar to the DESI LRGs (see Section~\ref{sec_GDD:results}). Solid lines show the sum of distribution function terms, calculated from the simulations, up to $\mu^6$. Dotted lines, only distinguishable at higher values of $\mu$, show the sum up to $\mu^4$.}
\label{fig:direct_vs_sum}
\end{figure}

In \cite{okumura12b}, these simulations were analyzed to extract the velocity moment power spectra for halos (see Section~\ref{sec:DF}). To obtain $P(k,\mu)$ for model comparison, we sum these terms following the distribution function expansion to include all contributions up to $\mu^6$. This provides greater resolution in $\mu$ than using $P(k,\mu)$ directly calculated in redshift space. Figure~\ref{fig:direct_vs_sum} compares $P(k,\mu)$ constructed with these two approaches, with the shot noise removed. They are in reasonable agreement, when compared with the expected measurement uncertainty, indicating that we can neglect higher powers of $\mu$ for $k \lesssim 0.3 \hMpcinv$.

\section{The distribution function approach}
\label{sec:DF}
The recently developed distribution function (DF) approach to redshift-space distortions expands the density field in redshift-space in terms of velocity moments of the distribution function \citep{seljak11}. The resulting redshift-space power spectrum is expressed in terms of real-space correlations between mass-weighted powers of the radial velocity field:
\begin{align}
\label{eq:DF_expansion}
P^{\rm s}(\mathbf{k})=\sum\limits_{L=0}^{\infty}\sum\limits_{L'=0}^{\infty}\frac{\left(-1\right)^{L'}}{L! L'!}\left( \frac{i k \mu}{\mathcal{H}}\right)^{L+L'}P^{\rm r}_{LL'}(\mathbf{k}) ,
\end{align}
where $\mathcal{H}=aH$, and $P^{\rm r}_{LL'}$ is the real-space power spectrum of density weighted powers $L$ and $L'$ of the radial velocity field. For instance, $P_{00}$ is the standard density auto-power spectrum, while $P_{01}$ is the cross-power spectrum of density and radial momentum.

This approach naturally produces an expansion of $P(k,\mu)$ in powers of $\mu$, which is convergent on sufficiently large scales, with a finite number of correlations contributing at a given power of $\mu$. Thus, when considering geometric distortions, we use an expansion in powers of $\mu$ rather than multipoles, which have contributions from all higher powers of $\mu$ and thus, in principle, an infinite number of correlations between velocity moments.

This expansion is valid for the density field of both dark matter and biased tracers such as halos or galaxies. The $P_{LL'}$ terms have recently been explored in simulations \citep{okumura12b,okumura12a} and calculated perturbatively \citep{vlah12,vlah13}. In this work, we are interested in using the DF expansion to provide a more accurate model that can be fit to observations of clustering in redshift-space. While N-body simulations can provide such a model for halos directly, spanning the necessary cosmological parameter space is not feasible.\footnote{Recent work on simulation-based cosmic emulators, e.g.\ \cite{kwan13}, may provide an alternative approach.} Instead, we seek an analytic (or hybrid) approach that allows the rapid calculation of the redshift-space clustering for a given cosmology.
\subsection{Halo biasing}
\label{sec:biasing}

A description of halo clustering requires additional parameters to account for the unknown relationship between biased tracers and the underlying density field. As discussed in Section~\ref{sec:formalism1}, we employ a non-linear and non-local biasing model, resulting in four bias parameters for each redshift and halo mass bin: $\{b_1,b_2^{00},b_2^{01},b_s\}$, where $P_{00}$ and $P_{01}$ have different values for quadratic, local bias $b_2$ (see \cite{vlah13} for further discussion). Although it is not included in this parametrization, the contribution from a non-local third-order bias, $b_3^{\rm NL}$ \citep{saito13inprep}, remains significant and is responsible for the two different values of $b_2$. We would have obtained similar results with a bias parametrization using $\{b_1,b_2,b_3^{\rm NL},b_s\}$. While the nonlinear bias values could each be treated as an independent parameter, doing so would ignore theoretical understanding of the relationship between them (e.g.\ \cite{baldauf12, chan12}) and would reduce the constraining power of the observations. Instead, we treat them as functions of the linear bias $b_1$, yielding one redshift-dependent bias parameter for each halo mass bin. To leading order, $b_s=(-2/7) (b_1-1)$. The $b_2$ parameters have an approximately quadratic dependence on $b_1$, which is fit to the simulations. The model also contains a halo velocity dispersion $\sigma_v$. The linear theory prediction of this term (e.g.\ \cite{vlah13}) is sufficiently accurate for the model: $\sigma_v \approx \sigma_{v,{\rm lin}} \sim f(z)D(z)$. Thus the velocity dispersion contributes no \mbox{additional free parameters.}
\subsection{Stochasticity}
\label{sec:stochasticity}

A further complication to modeling the clustering of halos is their stochastic nature as a tracer of the density field. This contribution is commonly modeled as a Poissonian shot noise: $P_{\rm shot} = 1/\bar{n}$, for mean number density $\bar{n}$. However scale-dependent corrections due to halo exclusion and non-linear clustering can be significant \citep{baldauf13}. Although our Fisher forecast assumes Poissonian shot noise when assessing the information content of the power spectrum, we must account for deviations from this simple assumption when comparing the clustering of halos in simulations (which has non-Poissonian shot noise) to analytic models.

Significant progress has been made in understanding these non-Poissonian contributions (e.g.\ \cite{baldauf13}). However, a complete and reliable model does not yet exist. Instead, the full stochastic term, $\Lambda(k)$, can be estimated following the approach of \cite{seljak09}:
\begin{align}
\label{eq:lambdak}
\Lambda(k) = P^{\rm hh}_{00}(k) - 2b_1 P^{\rm hm}_{00}(k) + b_1^2P^{\rm mm}_{00}(k),
\end{align}
where ``h'' and ``m'' refer to halos and matter, respectively, and correlations are calculated in real space. The most conservative treatment of shot noise would allow for marginalization over a multi-parameter model able to capture the relevant scale-dependence (e.g.\ Equation~2.23 of \cite{vlah13}). Such an approach would suppress information on small scales, where shot noise dominates. Conversely, a reliable model for shot noise would provide information on halo or galaxy properties from the additional scale-dependence. In this work, we take an intermediate approach. We assume that the non-Poissonian corrections are understood such that no additional marginalization is required. However, when performing forecasts, these terms are left out of the derivatives with respect to the parameters of interest, and thus the shot noise correction contributes no information.

\begin{figure}[h!]
\begin{center}
\resizebox{\hsize}{!}{
\includegraphics{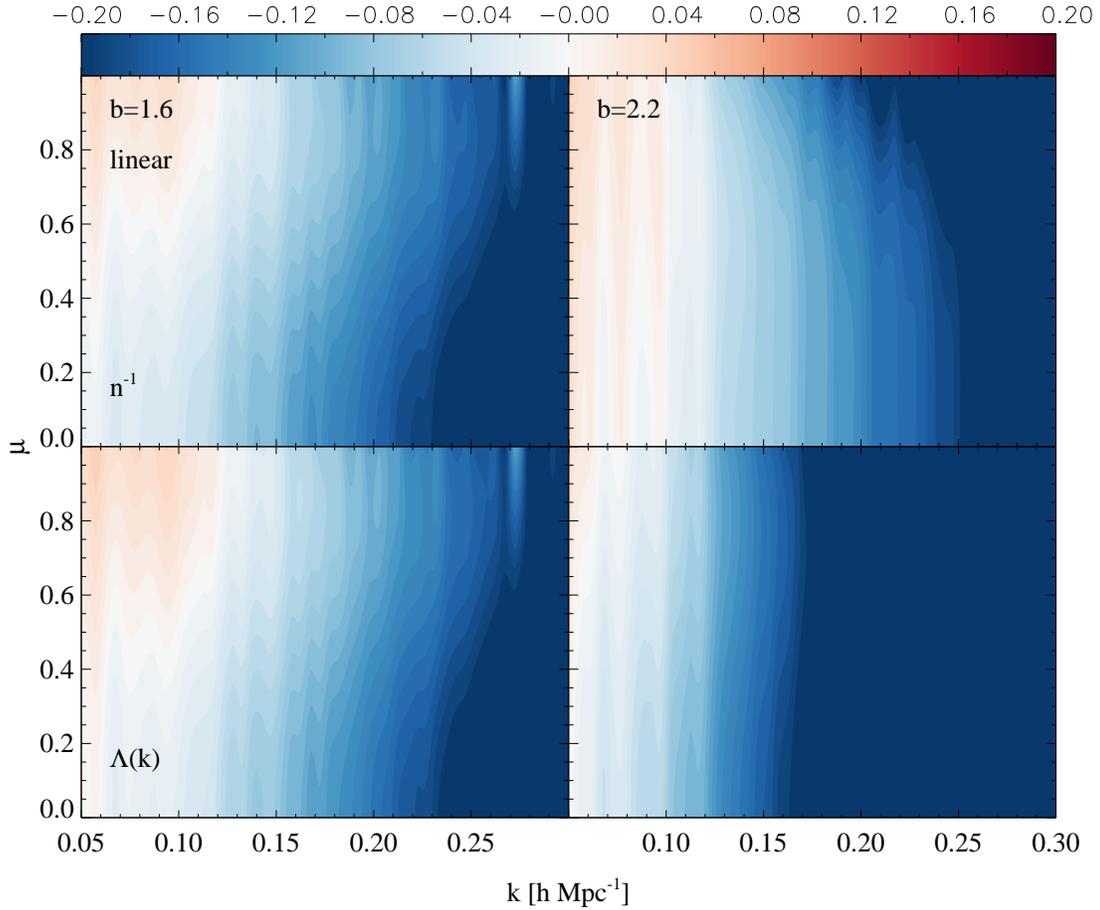}
}
\end{center}
\caption{The fractional difference ($P_{\rm model}/P_{\rm sim}-1$) between the linear Kaiser model and simulations is shown at $z=0.5$, for halos with $b_1=1.6$ ({\it left panels}) and $b_1=2.2$ ({\it right panels}). {\it Top panels:} Simulations have been corrected assuming the standard Poissonian shot noise. {\it Bottom panels:} The full $k$-dependent shot noise is used (as discussed in Section~\ref{sec:stochasticity}). The non-Poissonian correction is more significant for higher bias halos. Note that fractional differences are truncated at $\pm 0.20$.}
\label{fig:DF_model_resid1}
\end{figure}

\begin{figure}[h!]
\begin{center}
\resizebox{\hsize}{!}{
\includegraphics{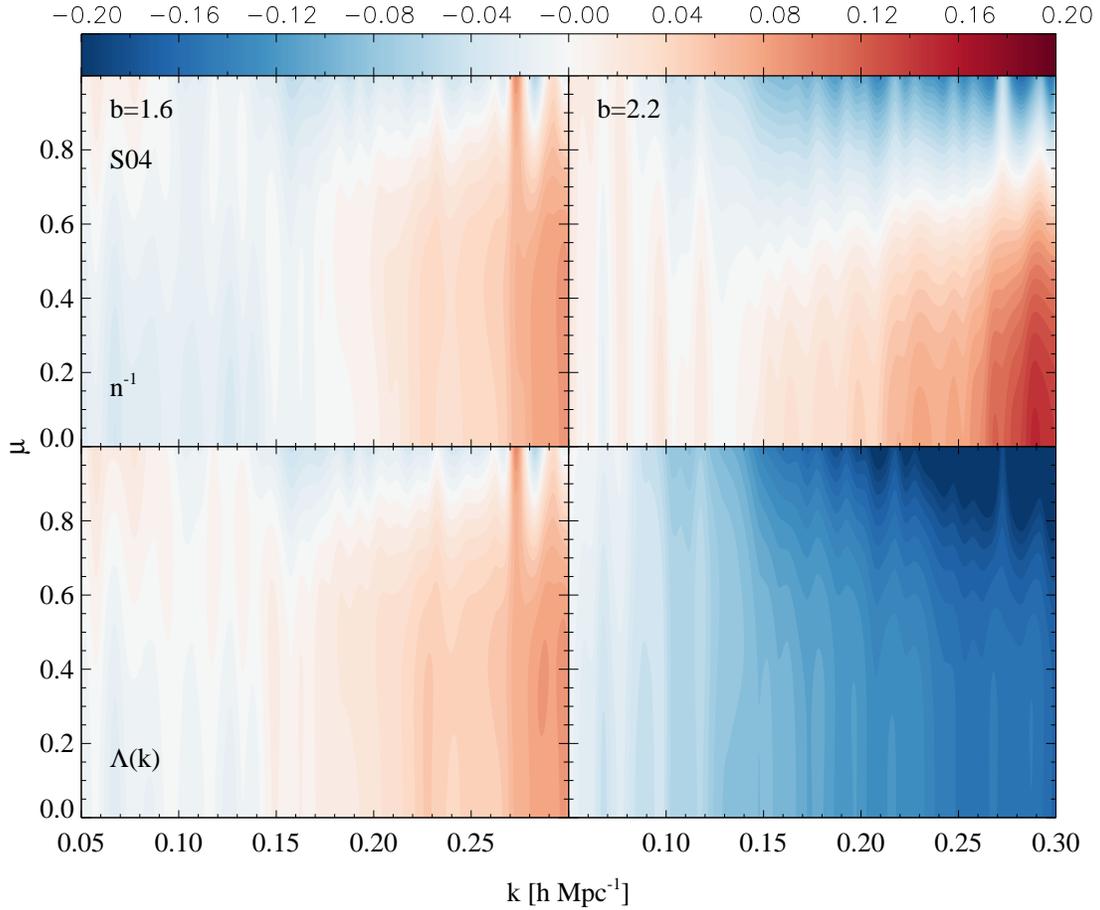}
}
\end{center}
\caption{The S04 model is compared to simulations, with the same conventions as Figure~\ref{fig:DF_model_resid1}.}
\label{fig:DF_model_resid2}
\end{figure}
\begin{figure}[h!]
\begin{center}
\resizebox{\hsize}{!}{
\includegraphics{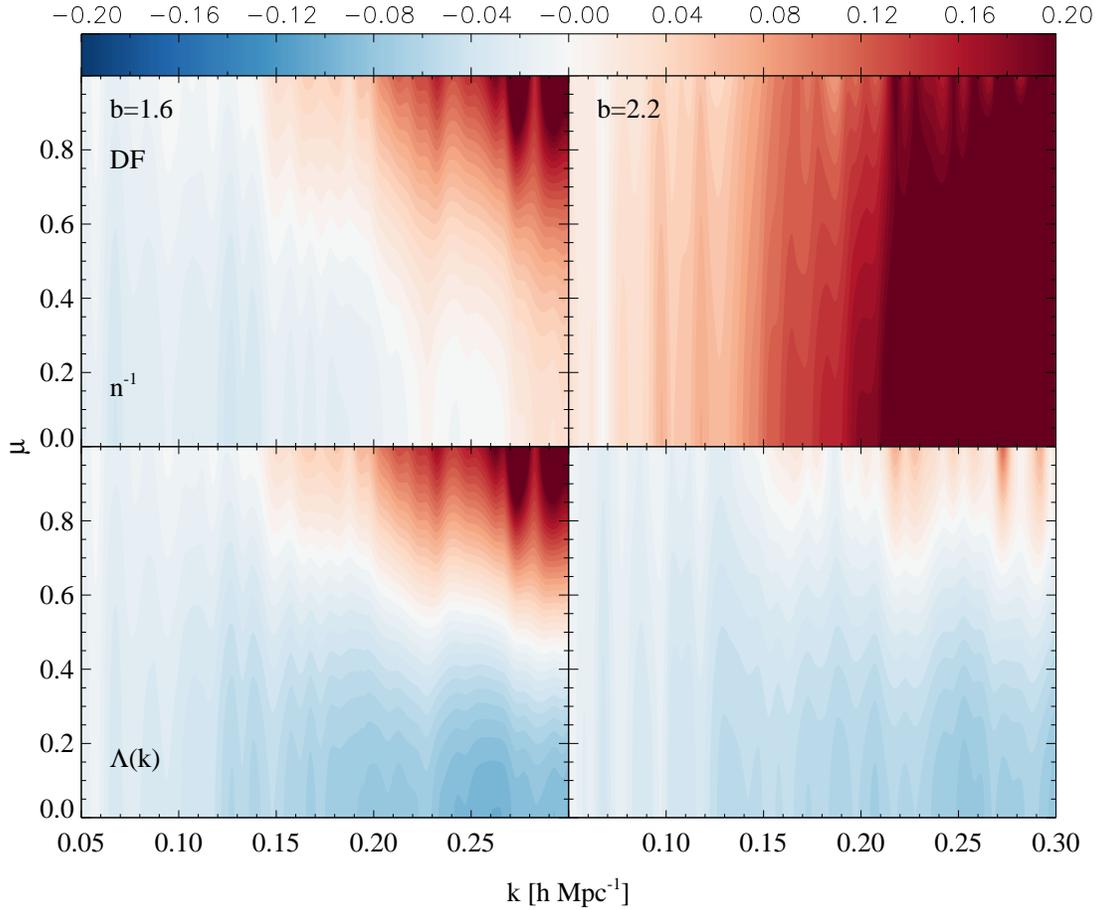}
}
\end{center}
\caption{The DF model is compared to simulations, with the same conventions as Figure~\ref{fig:DF_model_resid1}. The model includes the SPT corrections (see Section~\ref{sec:combining_terms}) and has $a_4=0$ (see Equation~\ref{eq:DF_model}).}
\label{fig:DF_model_resid3}
\end{figure}

\begin{figure}[h!]
\begin{center}
\resizebox{\hsize}{!}{
\includegraphics{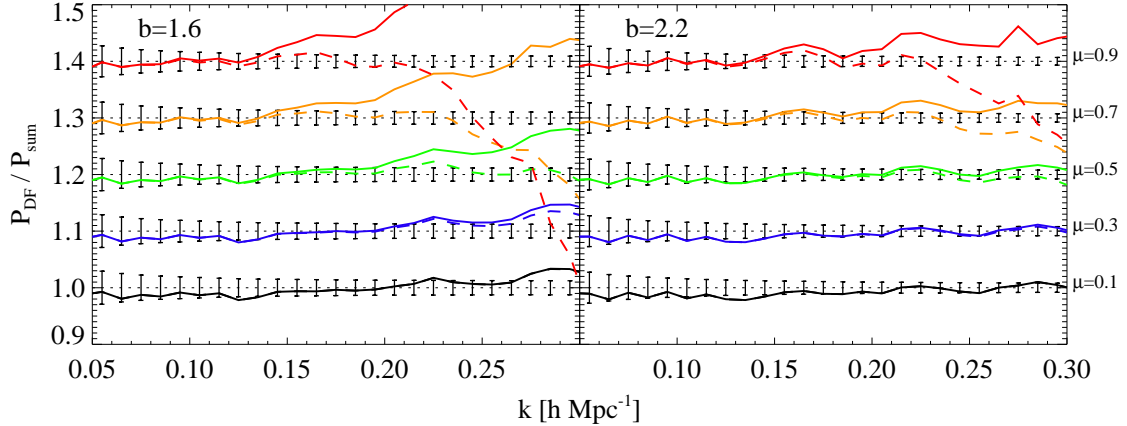}
}
\end{center}
\caption{The fractional difference between the distribution function model and the simulation $P(k,\mu)$ is shown for halos with bias 1.6 and 2.2 at $z=0.5$. Solid lines show the DF model with $a_4=0$. Dashed lines use the best-fit $a_4$ value, as discussed in Section~\ref{sec:model_corrections}. Five evenly spaced angular bins for $0 < \mu< 1$ are shown, with vertical offsets added for clarity and central $\mu$ values as labelled. Comparison is made with the summed $P(k,\mu)$ calculated using terms up to $\mu^6$. Error bars show the expected (fractional) uncertainty for a survey similar to the DESI LRGs. The $k$-dependent stochasticity is removed from the simulation results, and the SPT correction is included in the DF model.}
\label{fig:DF_model_sum}
\end{figure}
\begin{figure}[h!]
\begin{center}
\resizebox{\hsize}{!}{
\includegraphics{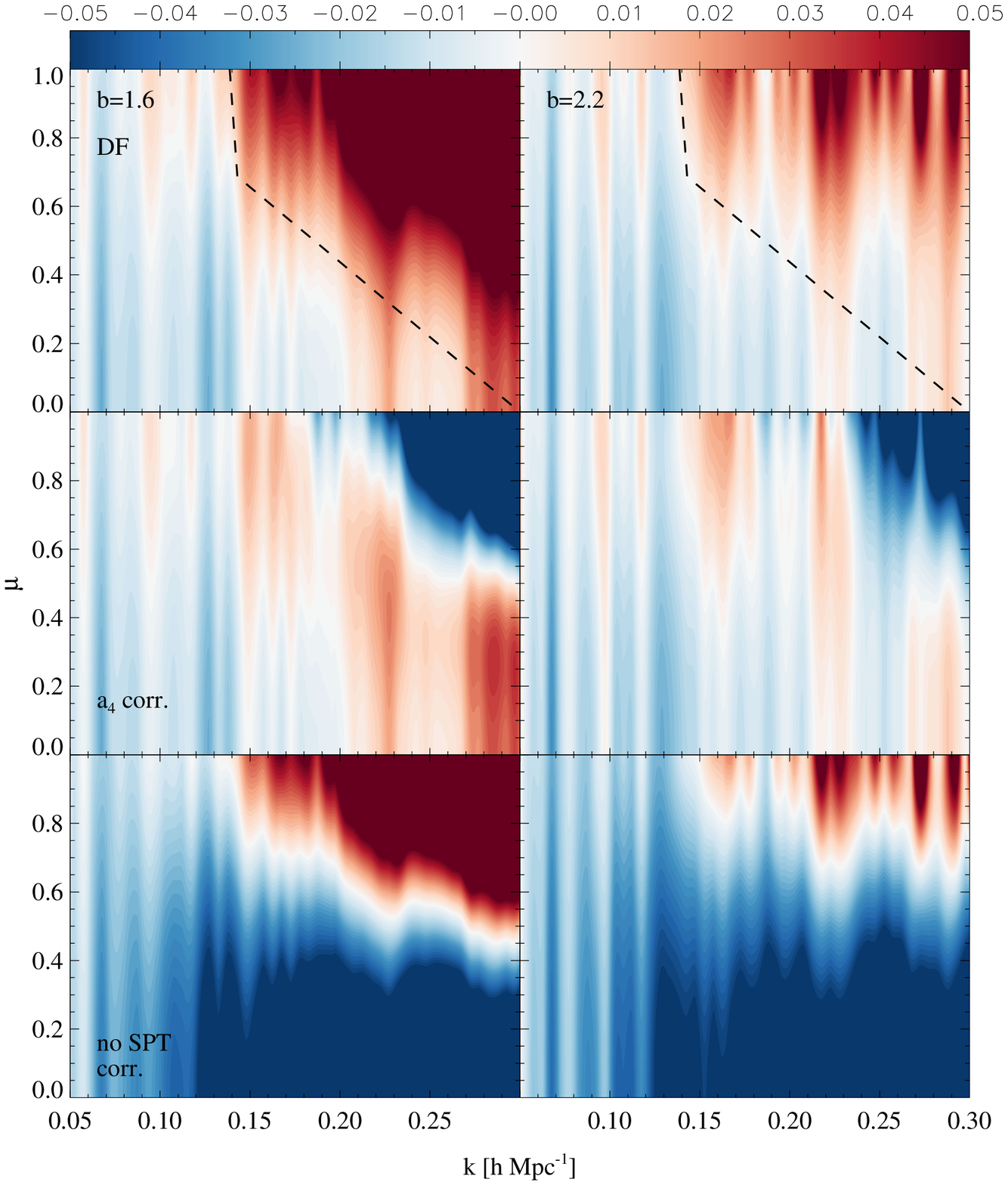}
}
\end{center}
\caption{The fractional difference between the distribution function model and simulations is shown at $z=0.5$. In all cases, the full $k$-dependent shot noise is used. {\it Top panels:} The DF model with $a_4=0$ and with the SPT corrections discussed in Section~\ref{sec:combining_terms}; {\it middle panels:} the DF model with the SPT corrections and best-fit $a_4$ correction (Equation~\ref{eq:DF_model}); {\it bottom panels:} the DF model with $a_4=0$ and without the SPT corrections. The dashed line indicates a $\mu$-dependent cut in $k$ that could increase the unbiased signal for parameter estimation, as discussed in Section~\ref{sec:optimal}. Note that the fractional difference is truncated at $\pm 5\%$ instead of $\pm 20\%$ as done in Figures~\ref{fig:DF_model_resid1}-\ref{fig:DF_model_resid3}.}
\label{fig:DF_model_resid4}
\end{figure}

\subsection{Combining terms}
\label{sec:combining_terms}
For simplicity, we choose to consider terms in the DF expansion (Equation~\ref{eq:DF_expansion}) up to $\mu^4$. Although including higher powers of $\mu$ would improve the model accuracy at $\mu \approx 1$, as seen in Figure~\ref{fig:direct_vs_sum}, these terms on the scales of interest tend to be smaller than the expected survey measurement uncertainty. Moreover, higher powers of $\mu$ have contributions from an increasing number of $P_{LL'}$ correlations and are thus more computationally intensive. We write the halo power spectrum in redshift space:
\begin{align}
\label{eq:DF_model}
P^{\rm hh}(k,\mu)=A(k)+B(k)\mu^2+(1+a_4 k^2)C(k)\mu^4,
\end{align}
where the $k$-dependent terms are determined by summing the relevant components of the real-space $P^{\rm hh}_{LL'}$ correlations:
\begin{align}
\label{eq:DF_model2}
A(k) &= P^{\rm hh}_{00,\mu^0}(k) ,\\
B(k) &= P^{\rm hh}_{01,\mu^2}(k) + P^{\rm hh}_{02,\mu^2}(k)+ P^{\rm hh}_{11,\mu^2}(k) , \notag \\
C(k) &= P^{\rm hh}_{11,\mu^4}(k) + P^{\rm hh}_{02,\mu^4}(k) + P^{\rm hh}_{12,\mu^4}(k) + P^{\rm hh}_{03,\mu^4}(k) + P^{\rm hh}_{13,\mu^4}(k) + P^{\rm hh}_{22,\mu^4}(k) + P^{\rm hh}_{04,\mu^4}(k) . \notag
\end{align}
As discussed in Section~\ref{sec:model_corrections}, the $a_4$ correction can be included to correct deficiencies in the biasing model as well as to partially account for missing higher powers of $\mu$.

The DF model terms are calculated analytically by applying the nonlinear biasing model to the relevant $P^{\rm hh}_{LL'}$ terms calculated with Eulerian perturbation theory, using a combination of standard perturbation theory (SPT) and resummation techniques (see \citep{vlah13}). However, the accuracy of the SPT calculations can break down on comparatively large scales. While recent and ongoing work has made large improvements in perturbative techniques (e.g.\ \cite{crocce06,pietroni08,taruya12}), we seek here to test the DF mapping from real to redshift space and the nonlinear biasing scheme, rather than the accuracy of a particular perturbation theory approach to describing the dynamics of dark matter. For some of the dominant terms in the model, where accurate SPT predictions are particularly challenging - namely the scalar parts of $P^{\rm mm}_{00}$, $P^{\rm mm}_{01}$, and $P^{\rm mm}_{11}$ - we apply a scale-dependent correction factor to the SPT terms to bring them into agreement with the dark matter results from N-body simulations. This correction is applied before determining the best-fit bias values. The discrepancy between SPT and dark matter simulations can be seen in Figures 1-3 of \cite{vlah12}, and its effect on the overall model of $P(k,\mu)$ is shown here in Figure~\ref{fig:DF_model_resid4}.

Once the shape of the linear power spectrum is fixed, the model can be expressed with the following parameters: $\{f(z),\sigma_8(z),b_1(z),a_4(z)\}$, in addition to the geometric distortions discussed in Section~\ref{sec:geo_dist}. The parameter $\sigma_8(z)$ refers here to the amplitude of the linear matter power spectrum at redshift $z$. Figures~\ref{fig:DF_model_resid1}-\ref{fig:DF_model_resid4} show the accuracy of the DF model, as well as the linear Kaiser model and the S04 model with no FoG (since the velocity dispersion for halos is small). Figure~\ref{fig:DF_model_resid1} compares the Kaiser model with the simulation results for both Poissonian and scale-dependent shot noise. Figures~\ref{fig:DF_model_resid2} and \ref{fig:DF_model_resid3} show the same for the S04 and DF models, respectively. Figure~\ref{fig:DF_model_sum} shows the accuracy of the DF model compared to the measurement precision of $P(k,\mu)$ expected from a survey similar to the DESI LRGs, while Figure~\ref{fig:DF_model_resid4} compares different versions of the DF model, including the $\mu^4$ correction discussed in Section~\ref{sec:model_corrections}, with simulations. In these comparisons, shot noise is subtracted from the simulation results rather than added to the model. This choice is somewhat arbitrary - the scale-dependent contribution could alternatively be considered part of the bias model. However, this convention matches the form of the Fisher matrix calculation, where shot noise is assumed to be Poissonian. Note the apparent accuracy of the S04 model when standard Poissonian shot noise is applied. Although it is a less sophisticated model, chance cancellations of neglected effects yield surprising agreement with the simulations, particularly for $b_1=1.6$ halos. Apparent features on the largest scales in these figures are due to scatter in the simulation measurements.
\subsection{Correcting higher-order angular dependence}
\label{sec:model_corrections}

The distribution function model in Equations~\ref{eq:DF_model}-\ref{eq:DF_model2} is complete up to terms scaling as $\mu^4$. Higher angular terms quickly become significant on small scales, and the model must thus include these terms for large values of $k_{\parallel}=k\mu$. In addition to these higher $\mu$ terms, the model prediction for the $\mu^4$ term itself diverges from the simulation results at high $k$, primarily due to a breakdown in the biasing model \cite{vlah13}.

One or more free parameters can improve the model accuracy on small scales and must be constrained from the observations at the cost of statistical power, effectively removing information from poorly modeled modes. In principle, the resulting increase in accuracy allows reliable extraction of information on smaller scales, where the number of modes rapidly increases. It is thus worth exploring whether such an approach improves constraints on the parameters of interest.

We choose to add a single free parameter, $a_4$: $F_4(k) \mu^4 \rightarrow (1+a_4 k^4) F_4(k) \mu^4$, which depends on $k_{\parallel} = k\mu$. However, this choice is somewhat arbitrary, and other reasonable corrections could be applied. The $a_4$ parameter partially accounts for deviations between the $\mu^4$ term in simulations and calculated analytically (due to imperfect modeling of halo bias), as well as the missing higher $\mu$ terms, although the latter effect is sub-dominant on scales of interest. The value of $a_4$ must be determined from the data, with a fiducial value determined using $\chi^2$ minimization with respect to the simulation results, weighting by the number of modes and signal-to-noise in each $\mathbf{k}$-bin (ignoring shot noise), up to $k=0.2 \hMpcinv$. This choice was made to prevent smaller scales that are unlikely to be used in cosmological analyses from dominating the fit.  Figures~\ref{fig:DF_model_sum} and \ref{fig:DF_model_resid4} demonstrate the effect of this additional parameter. We compare the resulting parameter constraints with and without $a_4$ below. Note that geometric distortions of the $a_4$-term are ignored when taking derivatives for the Fisher matrix, since its particular form should not contribute any cosmological information.




\section{Forecasting measurement precision and bias}
\label{sec:fisher}

We wish to determine how well a galaxy redshift survey can constrain cosmological physics by measuring geometric distortions ($D_A$ and $H$) and dynamical distortions ($f$ or $f\sigma_8$). The achieved precision and accuracy depend on the underlying model for clustering in redshift space. Attempting to use measurements on small scales where the model breaks down will introduce a systematic bias to parameter estimates, even while reducing the statistical errors. In general, a more complicated model (with additional free parameters) will reduce this systematic bias at the cost of statistical constraining power.


We employ the Fisher matrix formalism to forecast parameter constraints around a fiducial cosmology. We assume a fixed shape for the linear power spectrum (a reasonable approximation, given the precision from Planck measurements), and only allow the following parameters (or a subset thereof) to vary: $\{ H, D_A, f, \sigma_8, b_1, a_4\}$.  As discussed in \cite{reid12}, relaxing a hard prior on the linear power spectrum shape has no discernible effect on the ultimate constraints on $H$, $D_A$, and $f$. In general, however, it is important to note that the BAO feature constrains $D_A / r_{\rm s}$ and $H r_{\rm s}$, for sound horizon $r_{\rm s}$. The quantities constrained with broadband features will depend on the cosmological parameters that determine their shape and scale, while the AP and volume effects directly measure combinations of $D_A$ and $H$. A more complete analysis including changes in the underlying cosmological parameters would consistently account for these dependencies. The growth factor, $G(z)$, is completely degenerate with $\sigma_8$, and it is thus not considered as a separate parameter. For linear theory, there are only two independent combinations of the set $\{b_1,f,\sigma_8\}$, namely $\{b_1\sigma_8, f\sigma_8\}$.  However, the distribution function model depends on all three independently, as discussed below.


\subsection{Fisher matrix formalism}
The amount of information about the parameters ${p_i}$ contained in a set of observables with covariance matrix $C$ is given by the Fisher matrix:
\begin{align}
F_{ij} = \frac{1}{2}{\rm Tr}\left[C,_i C^{-1} C,_j C^{-1}\right],
\end{align}
where $C,_i\equiv \partial C / \partial p_i$. For the two-dimensional redshift-space power spectrum $P(k,\mu)$, measured in a galaxy survey, the Fisher matrix can be written \citep{tegmark97,seo03}:
\begin{align}
\label{eq:F_ij}
F_{ij}=\int_{k_{\rm min}}^{k_{\rm max}} \frac{2\pi k^2 dk}{(2\pi)^3}\int_0^1 d\mu\left( \frac{\partial \ln P(k,\mu)}{\partial p_i}\right)\left( \frac{\partial \ln P(k,\mu)}{\partial p_j}\right)V_{\rm eff}(k,\mu),
\end{align}
where the effective volume determines the signal-to-noise for $P(k,\mu)$:
\begin{align}
\label{eq:V_eff}
V_{\rm eff}(k,\mu)=V_{\rm s}\left[\frac{\bar{n} P(k,\mu)}{1+\bar{n} P(k,\mu)}\right]^2,
\end{align}
for survey parameters $V_{\rm s}$ (volume) and $\bar{n}$ (mean galaxy number density). Equation~\ref{eq:F_ij} assumes that measurements for each $\mathbf{k}$-mode are independent, with uncertainty given by $\sigma_P/P \propto V_{\rm eff}^{-1/2}$, which includes both sample variance and Poissonian shot noise $\bar{n}^{-1}$. On sufficiently small scales, this formula breaks down as power spectrum measurements on different scales become correlated. For a given measurement, the Fisher matrix provides the minimum statistical uncertainty on $p_i$, marginalized over all other parameters: $\sigma^2_i=\left(F^{-1}\right)_{ii}$. Independent measurements can be combined by adding their respective Fisher matrices.

This formalism can also be used to estimate the systematic bias on $p_i$ that comes from assuming an incorrect model, denoted $P'$, instead of the true $P$. Following \cite{taruya11}, the bias on $p_i$ is given by:
\begin{align}
\label{eq:Fisher_bias}
\Delta p_i &= - \sum\limits_j \left(F'^{-1}\right)_{ij}s_j, \\
s_j &= \int_{k_{\rm min}}^{k_{\rm max}} \frac{2\pi k^2 dk}{(2\pi)^3} \int_0^1 d\mu \left(\frac{P'(k,\mu)-P(k,\mu)}{P'(k,\mu)} \right) \left( \frac{\partial \ln P'(k,\mu)}{\partial p_j}\right) V'_{\rm eff}(k,\mu), \notag
\end{align}
where $F'_{ij}$ and $V'_{\rm eff}$ are calculated using $P'$. This expression is derived by expanding around the maximum of the likelihood and is thus only valid when the parameter bias (i.e.\ the shift away from the maximum) is small. However, we are only concerned with the case $\Delta p_i \lesssim \sigma_i$, where Equation~\ref{eq:Fisher_bias} remains applicable. Beyond this point, the parameter estimate has been biased beyond the level of the statistical uncertainty and is unreliable. As seen below, a parameter estimate can quickly become biased as information from smaller scales is included. The number of modes rapidly increases, driving down the statistical errors, while simultaneously the modeling of nonlinear clustering, redshift-space distortions, and biasing becomes inaccurate.

For the ``true'' $P(k,\mu)$, we use the halo measurements from N-body simulations. Although we assume simple Poissonian shot noise in the Fisher calculation, the true shot noise is somewhat non-Poissonian and $k$-dependent (as discussed in Section~\ref{sec:stochasticity}). We subtract this full shot noise from the simulation $P(k,\mu)$.

Recently, \cite{font13} parameterized the loss of information on small scales due to nonlinear evolution by applying an overall suppression to the linear theory information, with a suppression factor equivalent to that in Equation~\ref{fig:BAO_damping}. Instead, our approach directly measures the information content in the nonlinear model and uses the systematic bias to indicate where this information is no longer reliable.  In addition to applying a $\kmax$, determined by the systematic bias, we could relax some of the assumptions made in constructing the DF model, leading to additional free parameters that would serve to decrease the information content.

\subsection{Separating information from the BAO feature}
We wish to isolate information coming from the BAO feature, which provides a known physical scale, and the broadband shape, which lacks sharp features. We model the power spectrum as $P=P_{\rm BAO} + P_{\rm BB}$ and separate these components with a basis spline, which fits the smooth broadband shape of the power spectrum. This technique is similar to that employed in \cite{percival10}. We mitigate the challenge of fitting the rapidly changing power spectrum by first dividing by the linear baryon-free approximation of \cite{eisenstein98}. For the linear power spectrum, where both an analytic approximation and a basis spline model for $P_{\rm BB}$ are available, we find no appreciable difference in BAO information content in the two approaches. We note the importance of high resolution input power spectra to model the BAO information. If the power spectrum is sampled at insufficient resolution, the amplitude of numerical derivatives of the BAO feature is reduced, leading to a spurious reduction in the forecast parameter sensitivity.

In the following sections, where results are shown for specific values of $k_{\rm max}$, all BAO information is included, and $k_{\rm max}$ refers to the broadband information only.\footnote{In this work, ``all BAO information'' means that $k^{\rm BAO}_{\rm max}=0.4~\hMpcinv$, a somewhat arbitrary choice that has no impact on the results, given the BAO damping.} Where results are shown as a continuous function of $k_{\rm max}$, BAO and broadband information are both truncated at $k_{\rm max}$.

\subsection{BAO damping and reconstruction}
\label{sec:BAOrecon}
Nonlinear evolution damps the BAO feature on small scales. These effects are automatically included in the nonlinear treatment of the distribution function model. For linear theory, we use the formalism of \cite{eisenstein07,seo07} to account for the loss of information due to this damping:
\begin{align}
\label{fig:BAO_damping}
P_{\rm BAO, NL}(k,\mu) &=P_{\rm BAO, lin}(k,\mu) \exp\left[-\frac{1}{2}k^2\Sigma^2\left((1-\mu^2)+(1+f)^2\mu^2\right)\right], \\
\Sigma(z) & \approx 9.4 \left(\frac{\sigma_8(z=0)}{0.9}\right)\left(\frac{D(z)}{D(z=0)}\right) \hMpc, \notag
\end{align}
where $\Sigma(z)$ is the rms Lagrangian displacement, which is enhanced (in redshift space) along the line-of-sight by RSD, yielding the factor of $(1+f)$. Reconstruction of the density field \citep{eisenstein07b} partially restores this information by removing some of the nonlinear displacements of density tracers. In the framework of Fisher forecasts, reconstruction effectively reduces $\Sigma$, with ``standard'' reconstruction giving $\Sigma \rightarrow \Sigma/2$.

To include BAO reconstruction with the DF model, we evaluate the Fisher matrix using full-shape information and then add the Fisher matrix corresponding to the difference between the reconstructed and non-reconstructed BAO-only information in linear theory. While it is feasible to model reconstruction by increasing the amplitude of the BAO information isolated from the DF model, we believe the procedure is more robust and has been more thoroughly explored in linear theory. A more detailed examination of BAO information isolated from nonlinear models is left for future work. 

\subsection{Optimal modeling and analysis}
\label{sec:optimal}
Table~\ref{tab:models} summarizes the different approaches we consider for extracting information from clustering measurements in redshift space. For convenience, these models are numbered 1-6 and are subsequently referred to by number. BAO-only analysis is done with and without reconstruction (models 1 and 2, respectively). We then consider three possible approaches to including broadband information. In models 3 (no reconstruction) and 4 (with reconstruction), we use the DF model without the correction term to account for deviations at high $k_{\parallel}$ (discussed in Section~\ref{sec:model_corrections}), corresponding to $a_4=0$ in Equation~\ref{eq:DF_model}. Model 5 employs the same DF model (including reconstruction) but applies a $\mu$-dependent cut in $k$ to remove modes at high $k_{\parallel}$, where nonlinear redshift-space distortions are particularly challenging to model. In this work, we use the cut denoted by the dotted line in Figure~\ref{fig:DF_model_resid4} (analyzing only modes below the line). A reasonably straightforward cut can remove the most problematic modes. Although the simple cut used here was chosen by eye, a more sophisticated selection is feasible. In model 6, we use the DF model (with reconstruction) and apply the $a_4$ correction. Note that reconstruction refers only to BAO information - broadband information is assumed to be unchanged.

Models 5 and 6 should reduce the systematic bias introduced by broadband model inaccuracy, although specific cases can yield different behavior. However, this improvement comes at the cost of statistical constraining power: in model 5 modes are explicitly removed, while model 6 requires simultaneously fitting an additional parameter that is partially degenerate with the parameters of interest. Since these methods predominantly remove information from modes along the line-of-sight, constraints on $H$ are affected more than those on $D_A$. In model 6, if strong priors could be placed on $a_4$ (e.g.\ from simulations), this loss of statistical power could be mitigated.

In principle, techniques like those in models 5 and 6 will allow an accurate fit to observations on smaller scales. The resulting trade-off between the loss of modes at a given $k$ and a higher unbiased $\kmax$ must be considered for different scenarios. We note, however, that the forecast assumes a Gaussian covariance matrix, which will no longer hold on sufficiently small scales. The effect of non-Gaussianity is to reduce the information content of additional modes. As a result, approaches that improve the model fit on smaller scales are less powerful than the Gaussian case would indicate. Given similar Gaussian forecasts for unbiased parameter constraints, a less complex model or analysis technique, with a correspondingly lower $\kmax$, is likely preferred.

\begin{table}[t!]
\begin{center}
\begin{tabular}{c | cccc}
Model & BAO recon.& Broadband & $\mu$ cut & $a_4$ \\
\noalign{\hrule height 1pt}
1 & no & no & no & N/A \\
2 & standard & no & no & N/A \\
3 & no & DF & no & 0 \\
4 & standard & DF & no & 0 \\
5 & standard & DF & yes & 0 \\
6 & standard & DF & no & best fit \\
\end{tabular}
\end{center}
\caption{The different models considered in this work are outlined. For convenience, results for each model refer to the labels defined here. See text for descriptions of these model choices.}
\label{tab:models}
\end{table}

\section{Forecast results}
\label{sec_GDD:results}
We forecast statistical constraints and systematic biases for parameters of interest as a function of maximum wavenumber (minimum scale) $k_{\rm max}$. We assume a survey similar to the LRG portion of the proposed DESI experiment, with a volume of $10.5\ (h^{-1}{\rm Gpc})^3$, corresponding to $0.3<z<0.8$, and number density $\bar{n}=3\times10^{-4}\ (\hMpc)^{-3}$. Due to finite simulation snapshots, we focus on results at $z=0.5$, although this is somewhat lower than the effective redshift of the DESI LRG targets. We show results for halos with biases corresponding to the two lowest mass bins: $b_1=1.64$ and $2.18$, respectively, at $z=0.5$ (see Table~\ref{tab:halo}). Note that the expected LRG bias at this redshift is $b_1\approx 2.2$. These results are not intended to provide an exact forecast for the DESI experiment, but rather to demonstrate the potential for improvement and challenges for this type of next-generation redshift survey. Figures~\ref{fig:H_D_bands1}-\ref{fig:f_err_bias1} show forecasts for parameter constraints and systematic biases. Results for the relevant parameters at different $\kmax$, including the approximate best (unbiased) constraints, are given in Table~\ref{tab:constraints}. Constraints scale trivially as $V_{\rm s}^{1/2}$, with a more complicated scaling with number density, due to the changing relative importance of shot noise. The effect of changing the galaxy bias is more complicated, since both the total signal-to-noise and the relative importance of redshift-space distortions are altered. These effects are discussed below. Since the number density and bias considered here are similar to the CMASS sample in BOSS (e.g. \cite{anderson12,font13}), it is straightforward to compare constraints from each. The approximate volume of the CMASS sample for the completed BOSS survey is $V_{\rm s} \approx 3.5\ (h^{-1}{\rm Gpc})^3$, yielding parameter uncertainties larger than those shown here by a factor $\sim (10.5/3.5)^{1/2} = 1.7$. Results from intermediate data releases include only a fraction of the final volume and are additionally increased by $f_{\rm complete}^{-1/2}$. The true $\bar{n}(z)$ for the CMASS sample is not constant, dropping below $3\times10^{-4}\ (\hMpc)^{-3}$ at higher $z$, which yields a smaller effective volume and further degrades the constraints.

\begin{table}[t!]
\begin{center}
\begin{tabular}{c|cc|cc}
 \multicolumn{1}{c|}{Model} & $\sigma [D_A(z)]~[\%]$ & $\sigma [H(z)]~[\%]$& \multicolumn{2}{c}{$\sigma [f\sigma_8(z)]~[\%]$}\\
 & \multicolumn{2}{c|}{$\kmax=(0.10, 0.15, 0.20, 0.25)~[\hMpcinv]$} & (free geometry) & (fixed geometry) \\
\noalign{\hrule height 1pt}
\multicolumn{1}{c|}{$b_1=1.6$}&&&&\\
1 & {\bf 1.0} & {\bf 2.1} & - & - \\ 
2 & {\bf 0.6} & {\bf 1.0} & - & - \\
3 & 0.8, 0.8, 0.7, {\bf 0.7} & 1.5, {\bf 1.2}, 0.9, 0.8 & 3.7, {\bf 2.9}, 2.4, 2.0 & 2.2, {\bf 1.3}, 1.0, 0.7 \\
4 & 0.6, 0.5, 0.5, {\bf 0.5} & 0.9, {\bf 0.8}, 0.7, 0.6 & 3.0, {\bf 2.4}, 2.0, 1.7 & 2.2, {\bf 1.3}, 1.0, 0.7 \\
5 & 0.6, 0.5, 0.5, {\bf 0.5} & 0.9, 0.8, {\bf 0.8}, 0.8 & 3.0, 2.4, 2.2, {\bf 2.1} & 2.2, 1.4, 1.2, {\bf 1.2} \\
6 & 0.6, 0.5, 0.5, {\bf 0.5} & 1.0, 0.9, {\bf 0.8}, 0.7 &  3.3, 2.4, {\bf 2.1}, 1.9 & 2.9, 1.7, {\bf 1.2}, 0.9 \\
\noalign{\hrule height 1pt}
\multicolumn{1}{c|}{$b_1=2.2$}&&&&\\
1 & {\bf 0.8} & {\bf 1.9} & - & - \\ 
2 & {\bf 0.5} & {\bf 0.9} & - & - \\
3 & 0.7, 0.7, 0.7, {\bf 0.7} & 1.4, 1.2, 1.0, {\bf 0.9} &4.6, {\bf 3.7}, 3.0, 2.6 & 2.6, {\bf 1.6}, 1.1, 0.8 \\
4 & 0.5, 0.5, 0.5, {\bf 0.4} & 0.8, 0.8, 0.7, {\bf 0.7} & 3.5, {\bf 2.8}, 2.3, 2.1 &  2.6, {\bf 1.6}, 1.1, 0.8 \\
5 & 0.5, 0.5, 0.5, {\bf 0.5} & 0.8, 0.8, 0.8, {\bf 0.8} & 3.5, 2.8, 2.4, {\bf 2.3} & 2.6, 1.6, 1.4, {\bf 1.3} \\
6 & 0.5, 0.5, 0.5, {\bf 0.5} & 0.9, 0.8, 0.7, {\bf 0.7} & 3.9, 2.8, {\bf 2.4}, 2.1 & 3.4, 2.0, {\bf 1.4}, 1.0 \\
\end{tabular}
\end{center}
\caption{Fractional parameter constraints are shown for each model at different values of $k_{\rm max}$ for $b_1=1.6$ ({\it top}) and $b_1=2.2$ ({\it bottom}) halos at $z=0.5$. See Table~\ref{tab:models} for description of models. Assumed survey corresponds to DESI LRGs. Bold numbers indicate the approximate $\kmax$ that minimizes statistical uncertainty without introducing systematic bias at greater than $1\sigma$. To avoid relying on optimistic assumptions on small scales, $\kmax$ is chosen to not exceed $0.25 \hMpcinv$. Here, the choice of $k_{\rm max}$ affects only broadband information; all BAO information is included. BAO-only provides no constraints on $f \sigma_8$. Constraints on $H$ and $D_A$ involve marginalization over all relevant parameters, including $\sigma_8$. Constraints on $f \sigma_8$ are shown with $\sigma_8$ fixed and for both free and fixed geometry; both cases use the highest unbiased $\kmax$ for free geometry. See discussion in Section~\ref{sec:growth_constraints}.}
\label{tab:constraints}
\end{table}

\subsection{Constraints on geometry}
\label{sec:geometry_constraints}

\begin{figure}[h!]
\begin{center}
\resizebox{\hsize}{!}{
\includegraphics{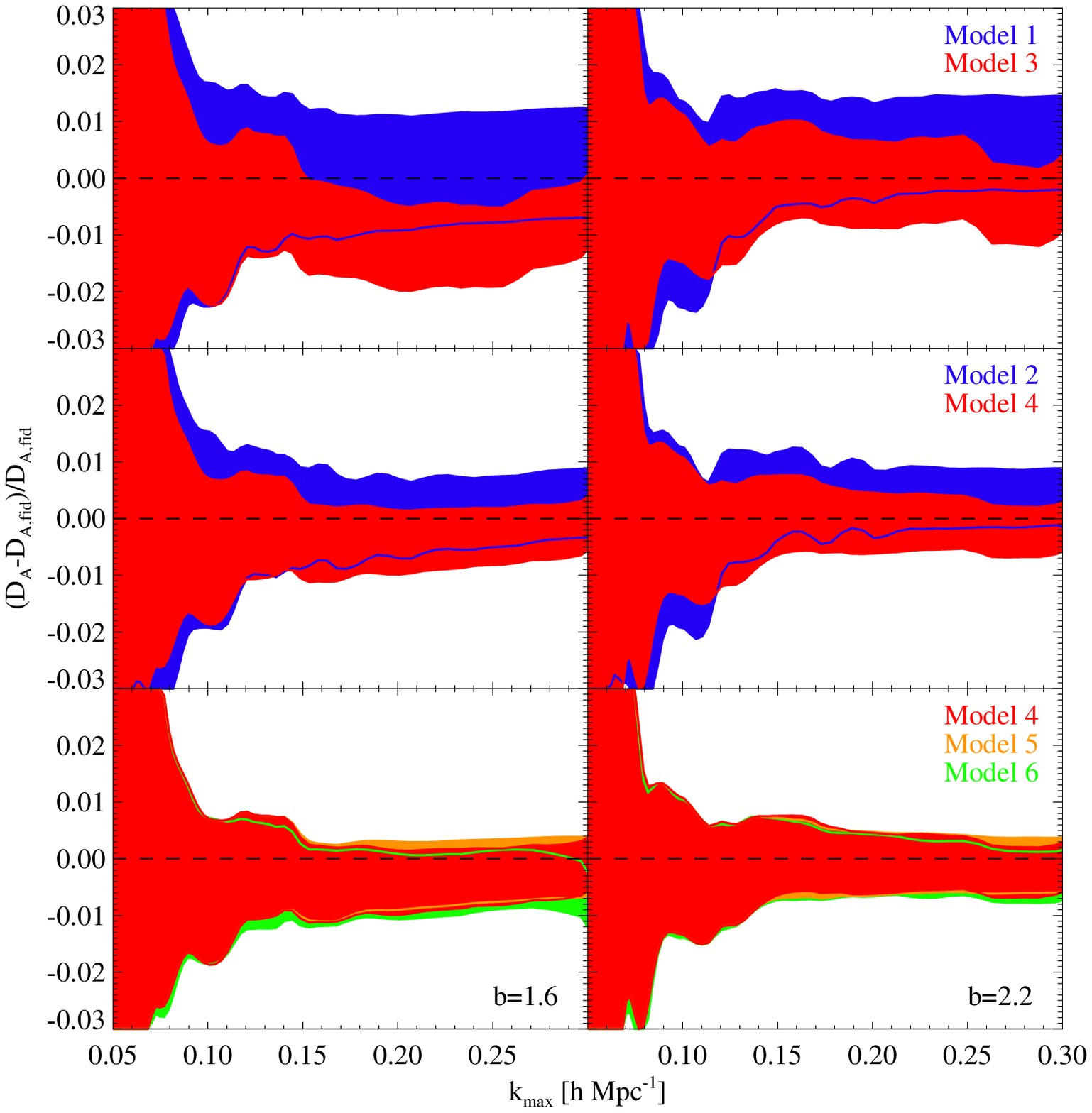}
}
\end{center}
\caption{Forecast for recovered $D_A(z)$ value and its statistical uncertainty ($1\sigma$), compared with the fiducial value, as a function of $k_{\rm max}$, for $b_1=1.6$ and $2.2$ halos at $z=0.5$. No information from BAO or broadband is included above the given $k_{\rm max}$. Assumed survey corresponds to DESI LRGs.
{\it Top panels:} Information from BAO-only (model 1; blue) is compared with full-shape information from the distribution function model (model 3; red).
{\it Middle panels:} Same as top, but standard BAO reconstruction is included: model 2 (blue); model 4 (red).
{\it Bottom panels:} Full shape information is compared for three different cases: model 4 (red); model 5 (yellow); model 6 (green). See Table~\ref{tab:models} for description of models.}
\label{fig:H_D_bands1}
\end{figure}

\begin{figure}[h!]
\begin{center}
\resizebox{\hsize}{!}{
\includegraphics{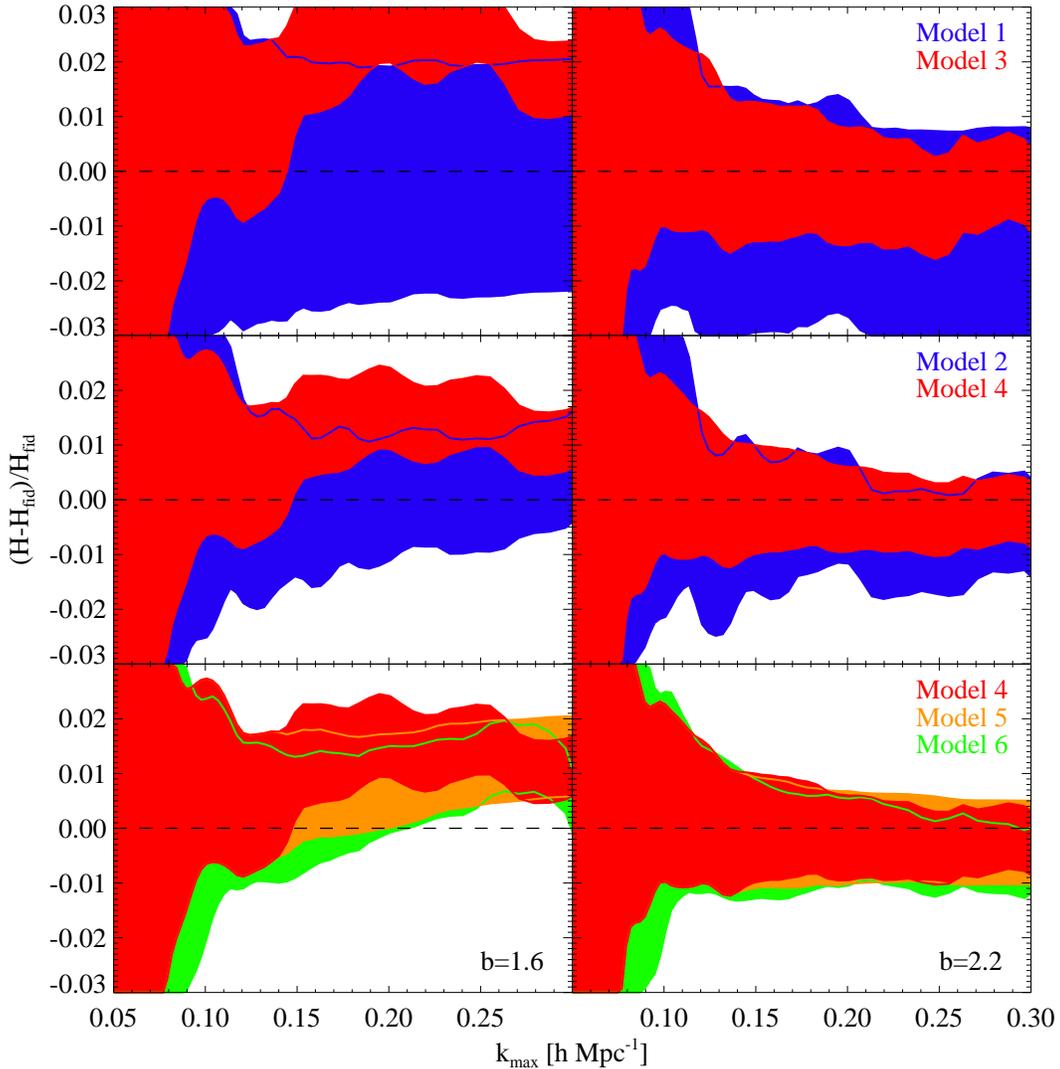}
}
\end{center}
\caption{Same as Figure~\ref{fig:H_D_bands1}, for $H(z)$.}
\label{fig:H_D_bands2}
\end{figure}

\begin{figure}[h!]
\begin{center}
\resizebox{\hsize}{!}{
\includegraphics{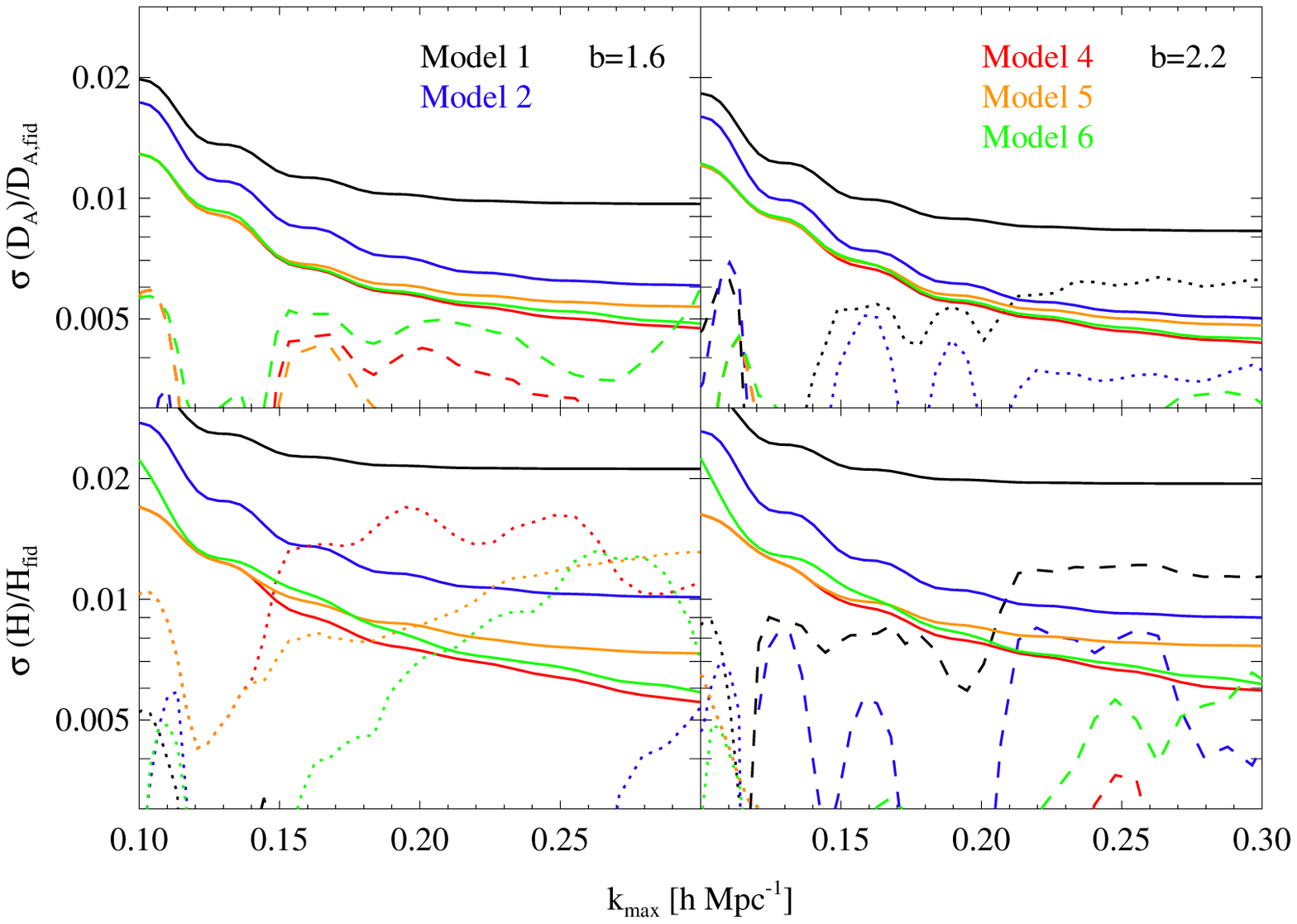}
}
\end{center}
\caption{Statistical precision (solid lines) and systematic bias on fractional measurements of $D_A$ ({\it top panels}) and $H$ ({\it bottom panels}) for $b_1=1.6$ and $b_1=2.2$ halos at $z=0.5$. Assumed survey corresponds to DESI LRGs. Positive (negative) systematic bias is indicated with dotted (dashed) lines. Model 1 (black); model 2 (blue); model 4 (red); model 5 (yellow); model 6 (green).}
\label{fig:H_D_err_bias1}
\end{figure}

Forecasts for measurements of geometric parameters $H$ and $D_A$, including both statistical precision and systematic bias, are shown in Figures~\ref{fig:H_D_bands1}-\ref{fig:H_D_err_bias1}. As expected, anisotropic BAO-only information is able to constrain geometry without introducing an appreciable systematic bias into the parameter results. Applying BAO reconstruction significantly improves the resulting constraints (by $\sim 50\%$ for $H$ and $\sim 40\%$ for $D_A$). These results are consistent with previous forecasts for anisotropic BAO information (e.g.\ \cite{seo07,font13}). Reconstruction provides a larger fractional improvement to constraints on $H$ than to those for $D_A$ because of the additional factor of $(1+f)$ suppressing the BAO in modes along the line-of-sight, which primarily contain information on $H$.


\begin{figure}[h!]
\begin{center}
\resizebox{\hsize}{!}{
\includegraphics{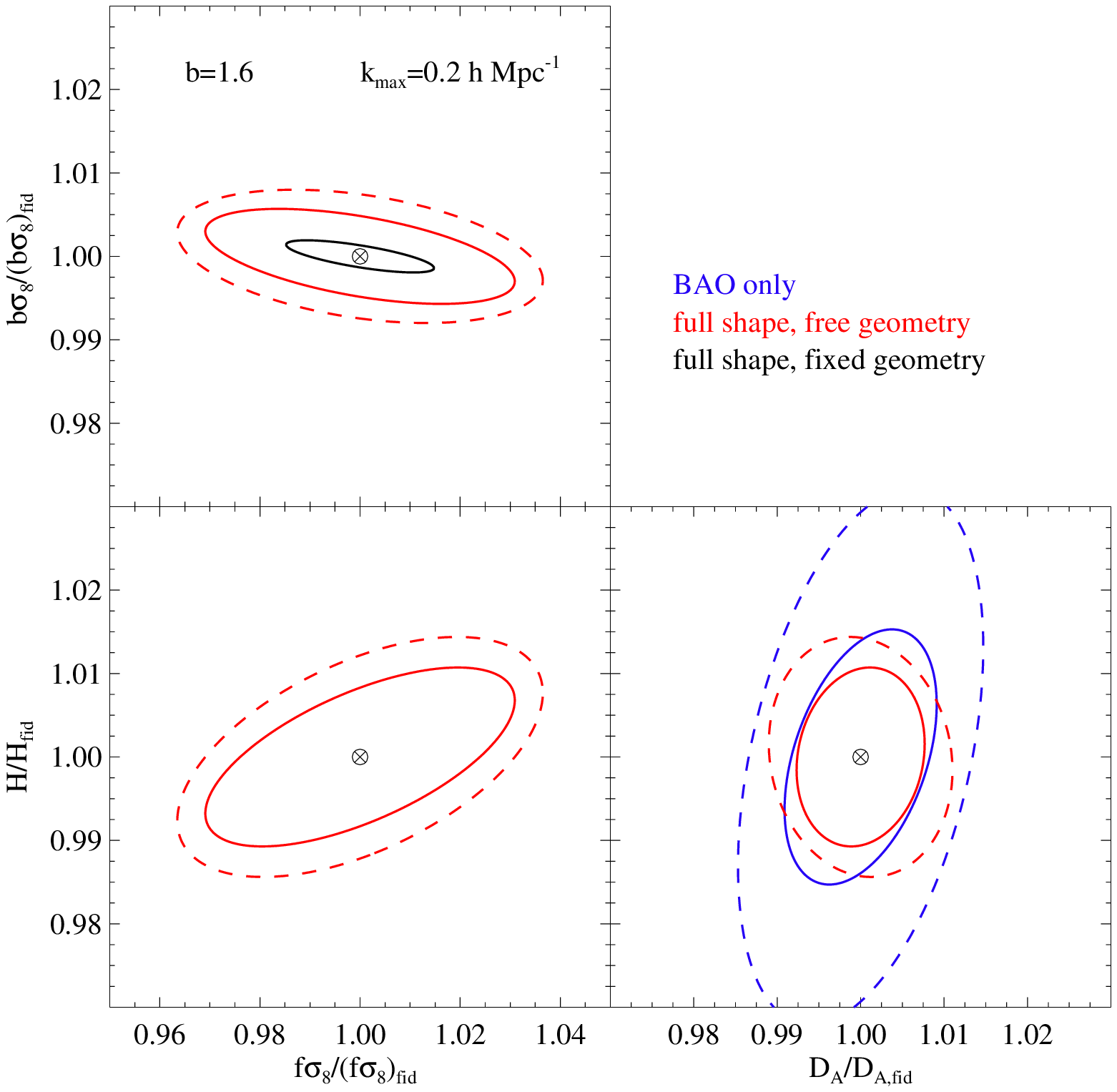}
}
\end{center}
\caption{Forecasts for 68\% joint statistical uncertainty regions are shown for $b_1=1.6$ halos at $z=0.5$. Assumed survey corresponds to DESI LRGs. Systematic parameter bias is ignored. Red and black lines indicate full-shape information has been used up to $k_{\rm max}=0.20~\hMpcinv$, with black lines indicating that geometry is held fixed. Blue lines in the $H-D_A$ plane correspond to BAO-only information. In all cases, all BAO information is included (i.e.\  beyond $k_{\rm max}$). Solid lines indicate that standard BAO reconstruction has been applied, while dashed lines have no reconstruction. As discussed in the text, $\sigma_8$ is held fixed to determine constraints on $f\sigma_8$ and $b\sigma_8$. For consistency, $\sigma_8$ is held fixed for constraints in the $D_A-H$ plane as well.}
\label{fig:ellipses1}
\end{figure}

\begin{figure}[h!]
\begin{center}
\resizebox{\hsize}{!}{
\includegraphics{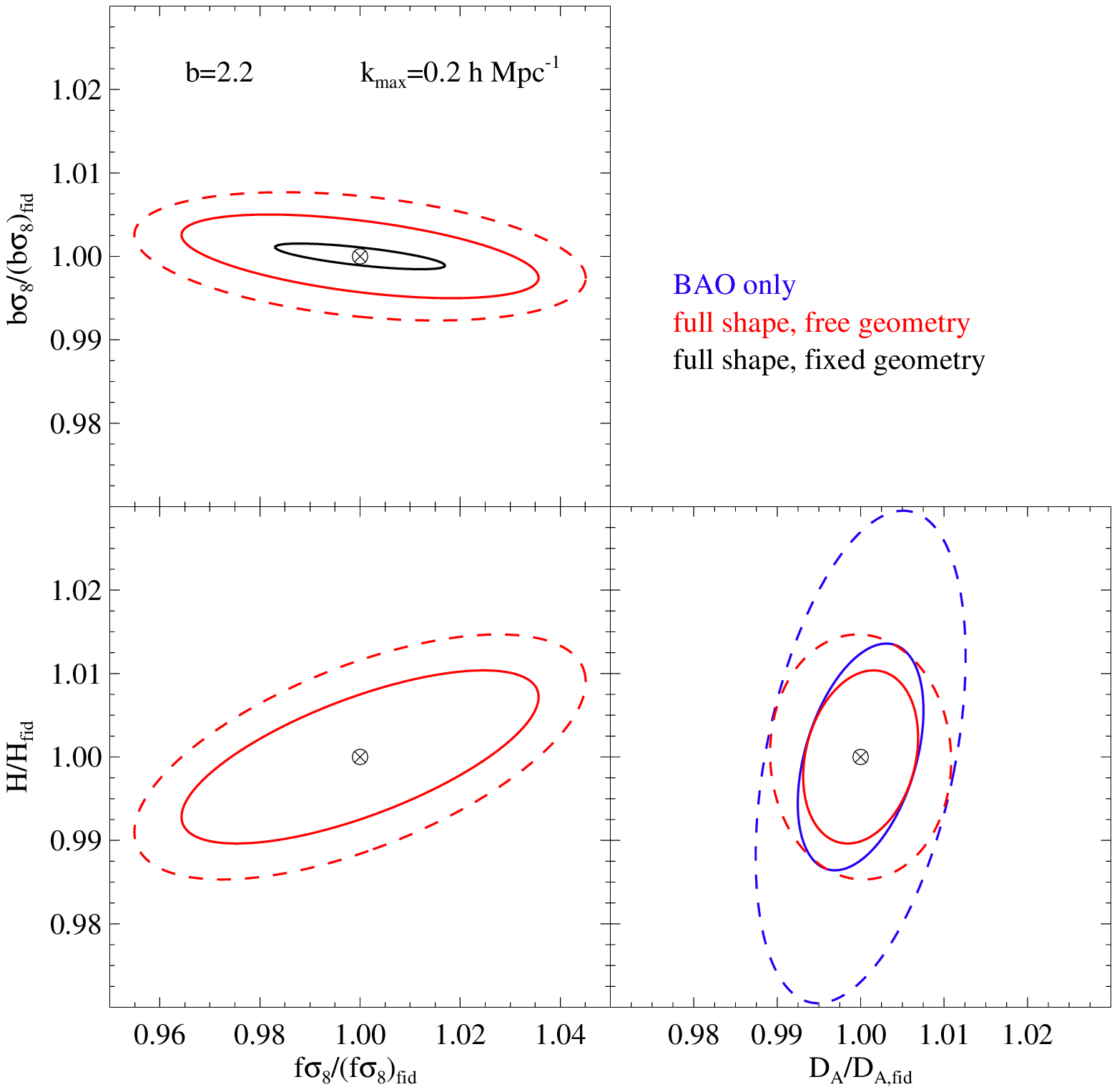}
}
\end{center}
\caption{Same as Figure~\ref{fig:ellipses1}, for $b_1=2.2$ halos.}
\label{fig:ellipses2}
\end{figure}

Including broadband information also improves parameter constraints. For $H$, including broadband information in the DF model without BAO reconstruction improves constraints by roughly the same amount as reconstruction would have in the BAO-only case ($\sim 50\%$). However, as seen in Figures~\ref{fig:ellipses1} and \ref{fig:ellipses2}, the degeneracy between $H$ and $D_A$ is significantly different for broadband information, which primarily provides an AP test. With reconstruction, including broadband information improves constraints on $H$ by an additional $\sim 30\%$. The improvement on $D_A$ is somewhat more modest, $\sim 20\%$ and $\sim 30\%$ with and without reconstruction, respectively. The reason for this disparity is discussed in Section \ref{sec:discussion}. The improvement from including broadband information depends strongly on the $\kmax$ which can be reliably used. In the sample-variance limited regime ($\bar{n}P \gg 1$), the square of the total signal-to-noise, corresponding to the amplitude of the Fisher matrix, should scale as the total number of modes ($\sim k_{\rm max}^3$), with corresponding parameter uncertainties decreasing as $\sigma \sim k_{\rm max}^{-3/2}$. In the shot-noise-dominated regime ($\bar{n}P \ll 1$), the additional information from including smaller scales decreases, with rapid saturation in the case of a steeply decreasing $P(k)$ (e.g.\ the linear theory prediction of $P(k) \propto k^{-3}$ on small scales). Nonlinear evolution yields an excess of power above linear theory, which both delays the onset of the shot-noise-dominated regime and slows the saturation of information once there. For $0.2 \lesssim k_{\rm max} \lesssim 0.3 \hMpcinv$, we find that nonlinear evolution leads to the Fisher matrix amplitude increasing roughly as $k_{\rm max}$, with parameter uncertainties thus decreasing as $\sigma \sim  k_{\rm max}^{-1/2}$. The results for models 4-6, seen in Figure~\ref{fig:H_D_err_bias1}, demonstrate this qualitative behavior. These approximate scalings are complicated by changing parameter degeneracies or a $\mu$-dependent $k$ cut.

For $b_1=1.6$ halos, systematic bias becomes an issue at $\kmax \approx 0.15 \hMpcinv$. As seen in Figure~\ref{fig:DF_model_resid4}, the DF model provides a somewhat better fit for the $b_1=2.2$ halos. The model is sufficiently accurate to avoid significant systematic bias of the results on scales as small as $\kmax=0.3 \hMpcinv$. However, this strong agreement may be partly due to optimistic assumptions regarding the ability to model non-linear biasing and non-Poissonian shot noise. We thus don't consider broadband information beyond $\kmax=0.25 \hMpcinv$.

Although the survey parameters used here (including redshift, volume, and number \mbox{density}) are chosen to approximate the LRG sample in DESI, it is instructive to compare parameter measurement precision from halos with different bias, with all other survey \mbox{characteristics} held fixed. In a simple model for $P(k,\mu)$, such as linear theory, low bias tracers provide a more powerful probe of $f\sigma_8$, as discussed in Section~\ref{sec:growth_constraints}. Because of the strong degeneracy between $H$ and $f\sigma_8$ (both affect line-of-sight modes), one would expect low bias tracers to provide better constraints on $H$ in the sample-variance regime. Indeed, forecasts for linear theory show this behavior, with the lower bias tracers outperforming higher bias tracers on $H$ until shot noise dominates, while for $D_A$, the higher bias tracers always provide tighter constraints. In the DF model, where nonlinear effects alter the signals and degeneracies for different parameters, these trends are not as clear. To illustrate the degeneracies between parameters, Figures~\ref{fig:ellipses1} and \ref{fig:ellipses2} show joint statistical constraints (without systematic bias) in the $H-D_A$, $H-f\sigma_8$, and $f\sigma_8-b_1\sigma_8$ planes.


The geometric signal coming from $\Delta V$, which changes the overall normalization of $P(k,\mu)$, is in principle quite large. However, in the case of linear theory, this effect is completely degenerate with $\{b_1\sigma_8,f\sigma_8\}$. In the distribution function model, $\Delta V$ is not completely degenerate with the parameter set $\{b_1,f,\sigma_8\}$, and thus it results in a modest change to forecast parameter constraints. This effect is larger for higher bias objects where nonlinear bias and matter clustering lead to less degeneracy between $\Delta V$ and other parameters.

Forecasts for $H$ and $D_A$ that include broadband information marginalize over $\sigma_8$ in addition to $b_1$ and $f$ (except as noted in Figures~\ref{fig:ellipses1}-\ref{fig:ellipses2}). Because $\sigma_8$ is largely degenerate with the other two (in combination), the additional marginalization has little impact on statistical uncertainties. However, because there is not total degeneracy between these parameters, holding $\sigma_8$ fixed can lead to a systematic bias in $H$ and $D_A$, since the modeling error can lead to a change in the preferred value of $H$ and $D_A$. The size and direction of this effect depends on the particular halo mass bin and minimum scale being considered.

Finally, the two more sophisticated analysis techniques explored here - applying a $\mu$-dependent cut in $k$ (model 5) and fitting an additional parameter to correct for higher $\mu$ dependence (model 6), do not yield improved overall constraints. The loss in statistical information is not sufficiently offset by a reduction in systematic bias.


\subsection{Constraints on growth of structure}
\label{sec:growth_constraints}
\begin{figure}[hbt]
\begin{center}
\resizebox{\hsize}{!}{
\includegraphics{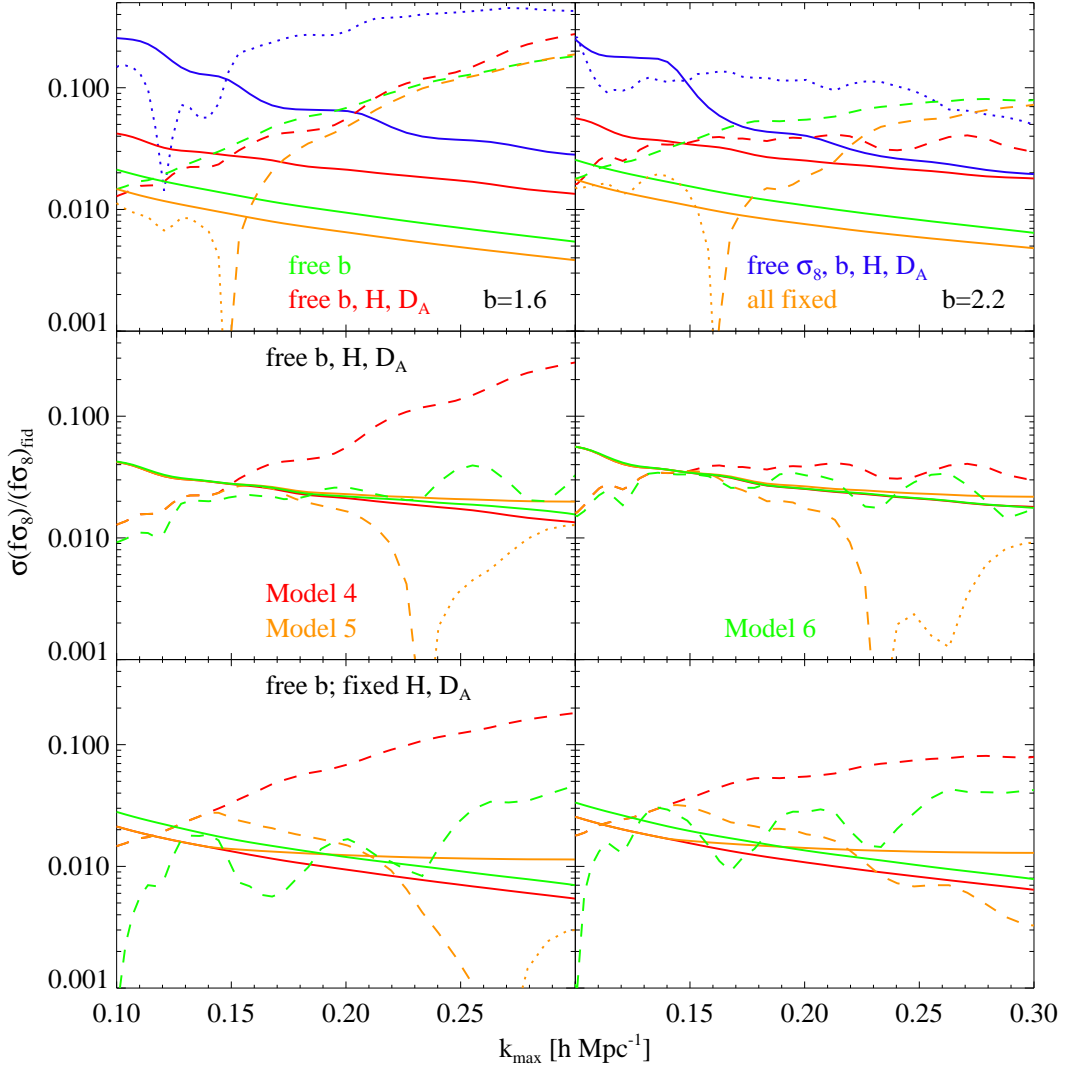}
}
\end{center}
\caption{Statistical precision (solid lines) and systematic bias on $f\sigma_8$ for $b_1=1.6$ and $b_1=2.2$ halos at $z=0.5$. Assumed survey corresponds to DESI LRGs. Positive (negative) systematic bias is indicated with dotted (dashed) lines. {\it Top panels:} The effect of allowing different parameters to vary in model 4 is shown. Blue line shows fractional precision and bias on $f$ alone (all other parameters, including $\sigma_8$, are allowed to vary). With $\sigma_8$ held fixed, thus leading to constraints on $f\sigma_8$: bias and geometry are allowed to vary (red); geometry is fixed (green); both bias and geometry are fixed (yellow). {\it Middle panels:} Constraints on $f\sigma_8$ ($\sigma_8$ held fixed; geometry and bias free) are shown for model 4 (red); model 5 (yellow); model 6 (green). {\it Bottom panels:} Same as middle panels, but with $\sigma_8$ and geometry held fixed; bias free.}
\label{fig:f_err_bias1}
\end{figure}

Figures~\ref{fig:ellipses1} and \ref{fig:ellipses2} show the forecast measurement precision for $f \sigma_8$ and $b_1 \sigma_8$, using full shape information in the DF model. Figure~\ref{fig:f_err_bias1} shows both statistical precision and systematic bias for measurements of $f \sigma_8$ (or $f$ alone, if $\sigma_8$ is simultaneously measured) using different versions of the DF model. While the linear theory $P(k,\mu)$ depends only on $f\sigma_8$ and $b_1\sigma_8$, the distribution function model has nonlinear contributions that scale with additional powers of $\sigma_8$, and nonlinear bias provides nontrivial dependence on $b_1$, thus breaking this degeneracy. As seen in Figure~\ref{fig:f_err_bias1}, this breaking occurs on small scales. In these figures (unless otherwise noted), constraints on $\{f\sigma_8, b_1\sigma_8\}$ are actually constraints on $\{f,b_1\}$ with $\sigma_8$ held fixed. In the limit where the $\sigma_8$ dependence appears only in $f\sigma_8$ and $b_1\sigma_8$ (e.g.\ in linear theory or on large scales in the DF model), these approaches are equivalent. As this degeneracy is broken, the two are no longer the same. However, the degeneracy is sufficient on the scales considered here that the deviation remains small. Despite differences in model and technique, we find constraints on $f\sigma_8$ (fixed geometry) that are consistent with the recent results of \cite{font13}.

As is apparent in Figures \ref{fig:ellipses1}-\ref{fig:f_err_bias1}, the degeneracy between geometry and growth of structure is quite strong. Indeed, at leading order, geometric and redshift-space distortions have the same effect on the power spectrum, introducing a $\mu^2$ dependence. Thus, strong priors on either geometry or growth of structure will allow significantly improved constraints on the other. These figures include the limiting case where geometry is fixed. Since linear redshift-space distortions have an amplitude characterized by $\beta \equiv f/b_1$, lower bias tracers provide a higher signal-to-noise measurement of RSD. Moreover, when trying to simultaneously constrain growth rate and geometry, higher $\beta$ reduces the degeneracy between RSD and geometric distortions by increasing the relative importance of higher $\mu$-dependence. These advantages are diminished at high $k$, where shot noise becomes significant. Similarly, although the BAO alone provides no constraints on $f \sigma_8$, it improves broadband-only constraints because it is able to separately measure $H$ and $D_A$, breaking the degeneracy with $f \sigma_8$, with further improvements from reconstruction.

Also of interest is the potential benefit of a more sophisticated model (using the $a_4$ correction) or analysis technique (applying a $\mu$-dependent cut in $k$). While these techniques seem to provide little advantage for geometric constraints, they can significantly improve the precision of an unbiased measurement of $f\sigma_8$ by allowing the inclusion of information at higher $k$ (see the middle and bottom panels of Figure~\ref{fig:f_err_bias1}). The improvement may be even more significant in the case where prior measurements allow geometry and/or bias to be held fixed. As is clear in the top panels of Figure~\ref{fig:f_err_bias1}, the systematic bias on $f\sigma_8$ is largely unaffected by whether or not geometry and bias are held fixed. Thus, even if separate measurements of geometry or bias existed, the uncorrected DF model (model 4) lacks sufficient broadband accuracy to take advantage of the improved statistical precision. Note that the optimal, unbiased constraints on $f\sigma_8$ for both free and fixed geometry, denoted as bold in Table~\ref{tab:constraints}, use the optimal $\kmax$ determined for free geometry.

\section{Discussion}
\label{sec:discussion}

The results in this work show the potential advantages of analyzing the broadband shape of $P(k,\mu)$, in addition to the BAO feature, when a sufficiently accurate model is used. Including broadband information when using linear theory provides little improvement: the systematic bias due to the inaccuracy of the model quickly overwhelms the statistical gain. Similar results have been seen in previous studies examining measurements of $f$ from broadband information (e.g.\ \cite{kwan12}), where model inaccuracy results in systematically biased parameter constraints on relatively large scales. As shown here, the situation can be significantly improved when using a more accurate model, such as the distribution function model of \cite{vlah13}. Applying this model, we demonstrate the potential for precision geometric measurements using broadband information, as well as improved constraints on growth of structure from measuring redshift-space distortions at smaller scales.

Including broadband information improves constraints on $H$ more significantly than those on $D_A$. This disparity is due to the different geometric degeneracies in BAO and broadband measurements. While the BAO feature is able to measure $H$ and $D_A$ separately, the primary signal is in the position of the angle-averaged BAO position, which roughly measures the parameter combination $D_A^2/H$. Thus, constraints are stronger on $D_A$. Equivalently, we can see that since there are more transverse modes than line-sight-modes, measurements of a transverse distance scale will be more precise. While similar arguments hold for any features in the broadband shape, the bulk of broadband information is coming from the Alcock-Paczynski test, which carries signal even in the absence of features at a known scale. Since this test measures $D_A H$ (the ``warping'' mode), it serves to break the degeneracy remaining in the BAO-only information, primarily benefiting the previously poorer constraints on $H$.
\FloatBarrier
Although there are differences in the models and techniques applied, our results are broadly consistent with other recent Fisher forecasts (e.g.\ \cite{seo07,font13}). However, in some cases, recent measurements on real or mock data sets have yielded less precise constraints (e.g.\ \cite{anderson12}). While some of this disparity may be due to the inherently optimistic nature of Fisher forecasts (they provide a lower limit on statistical uncertainty), there may be other effects contributing as well, such as survey geometry.

Our results indicate the potential value of adding external measurements able to break degeneracies between geometry, growth of structure, and bias. For instance, constraints on $f \sigma_8$ are significantly improved if geometry is already known. Similarly, the effect of the geometric factor $\Delta V$ is to cause an overall rescaling of the power spectrum amplitude. This effect is largely degenerate with $b_1$ and $\sigma_8$, even when nonlinear corrections are included. However, if these parameters can be measured separately, the signal-to-noise for measuring $\Delta V$ (and thus constraining $D_V$) is very high.

We also highlight advantages of low bias tracers, which have a higher signal-to-noise for measuring RSD. Alternatively, tracers with an additional velocity bias that increases the relative strength of RSD will exhibit similar advantages. We note that the considerations discussed in this work are not limited to galaxy redshift surveys, but also apply to any three-dimensional tracer of the density field, including neutral hydrogen seen in the Lyman-$\alpha$ forest and the 21cm line. Although an unknown velocity bias would be degenerate with measurements of $f$ (in the absence of accurate modeling), it is easier to measure broadband geometric distortions in the presence of stronger RSD, which are less degenerate, especially with $H$. These arguments suggest that performing an AP test with anisotropic Lyman-$\alpha$ clustering is an interesting possibility. Also, the use of multiple, overlapping tracers to reduce sample variance (e.g.\ \cite{mcdonald09, hamaus12}) can in some circumstances improve constraints.

As described in Section~\ref{sec:combining_terms}, SPT does not yield sufficiently accurate predictions for the dark matter correlations in some terms, and we thus apply corrections to recover the scale-dependence found in simulations. While the need for these corrections is a weakness of SPT, the aim of this work is not to test different perturbative schemes for calculating dark matter correlations. Instead, we seek to test the accuracy of the distribution function expansion and nonlinear biasing model as well as the corresponding implications for cosmological analyses. Our results demonstrate that this DF framework can provide very accurate predictions when a sufficiently accurate set of dark matter correlations is used. Ongoing developments in analytic and numerical approaches to rapidly estimate these dark matter correlations (including perturbative techniques and cosmological emulators based on suites of simulations), should soon reduce the need for these corrections. Although formulated in Fourier space, the distribution function model is able to provide accurate predictions for the halo correlation function in configuration space, as seen in Figures 18-19 of \cite{vlah13}. In the future, it would be informative to perform a more detailed comparison of this approach with other recently developed models, some of which are naturally expressed in configuration space (see, e.g.\, \cite{reid11,carlson13,wang14} for recent work modeling the velocities of biased tracers in a Lagrangian framework).

The parameters utilized in the DF model - multiple bias parameters, scale-dependent shot noise, and, when relevant, the $a_4$ term - reflect our current understanding of nonlinear biasing in redshift space. Without an underlying model for biasing, an additional free parameter is required to maintain consistency for each new statistic that is considered (e.g.\ halo auto-correlation, halo-matter cross-correlation, and density-momentum cross-correlation \cite{okumura13}). The distribution function approach provides a systematic and physically-motivated way to introduce these parameters and fit them to potentially observable correlations. Although the assumptions made in this work on our current ability to model these bias and stochasticity terms are somewhat optimistic, recent progress suggests that a more complete model will predict these terms as a function of halo mass. Moving from a description of halos to one of galaxies, including satellite objects and virial motions within halos, will affect the value of these parameters and introduce further complexity. These issues are the subject of ongoing and future work.

\acknowledgments
We thank Beth Reid, Hee-Jong Seo, Florian Beutler, Chris Hirata, and David Weinberg for useful discussions. We also thank an anonymous referee for helpful suggestions. J.B. appreciates the hospitality of the Institute for Theoretical Physics at the University of Zurich, where part of this work was done. This work is supported by the DOE, the Swiss National Foundation under contract 200021-116696/1, WCU grant R32-10130, and Ewha University research fund 1-2008-2935-001-2.



\providecommand{\href}[2]{#2}\begingroup\raggedright\endgroup


\end{document}